\documentclass[preprint,3p]{elsarticle}
\usepackage{amsmath} 
\usepackage{amssymb} 
\usepackage{booktabs}
\usepackage{colortbl}
\usepackage{enumitem}
\usepackage{graphicx}
\usepackage{subcaption}
\usepackage{rotating}
\usepackage[hyphens]{url}
\usepackage{hyperref}
\usepackage[normalem]{ulem}
\usepackage{xcolor}
\usepackage{xspace}
\definecolor{darkred}{rgb}{0.75,0.0,0.0}
\definecolor{darkgreen}{rgb}{0.0,0.6,0.0}
\definecolor{darkblue}{rgb}{0.0,0.0,0.6}
\definecolor{darkcyan}{rgb}{0.0,0.6,0.6}
\definecolor{darkmagenta}{rgb}{0.6,0.0,0.6}
\definecolor{darkamber}{rgb}{1.0,0.5,0.0}
\definecolor{darkyellow}{rgb}{0.6,0.6,0.0}

\definecolor{lightred}{rgb}{1.0,0.9,0.9}
\definecolor{lightgreen}{rgb}{0.9,1.0,0.9}
\definecolor{lightblue}{rgb}{0.9,0.9,1.0}
\definecolor{lightcyan}{rgb}{0.8,1.0,1.0}
\definecolor{lightmagenta}{rgb}{1.0,0.8,1.0}
\definecolor{lightamber}{rgb}{1.0,0.8,0.0}
\definecolor{lightyellow}{rgb}{1.0,1.0,0.8}

\definecolor{webgreen}{rgb}{0,0.5,0}
\definecolor{webbrown}{rgb}{0.6,0,0}

\definecolor{grey}{rgb}{0.65,0.65,0.65}
\definecolor{purple}{rgb}{0.4,0,0.75}

\definecolor{burgundy}{rgb}{0.5, 0.0, 0.13}             % For states
\definecolor{darkcyan}{rgb}{0.0,0.6,0.6}                % For sync messages
\definecolor{darkpastelgreen}{rgb}{0.01, 0.75, 0.24}    % For guarded transitions

\newcommand{\mynote}[2]{
	\ifstrequal{#1}{0}{\textcolor{darkamber}{#2}}{}%
  	\ifstrequal{#1}{1}{\textcolor{darkmagenta}{#2}}{}%
  	\ifstrequal{#1}{2}{\textcolor{darkcyan}{#2}}{}%
  	\ifstrequal{#1}{3}{\textcolor{darkgreen}{#2}}{}%
  	\ifstrequal{#1}{5}{\textcolor{darkblue}{#2}}{}%
  	\ifstrequal{#1}{8}{\textcolor{burgundy}{#2}}{}
}
\newcommand{\todiscuss}[2]{
	\ifstrequal{#1}{0}{\textcolor{darkamber}{\textit{\textbf{TO DISCUSS}: #2}}}{}%
  	\ifstrequal{#1}{1}{\textcolor{darkmagenta}{\textit{\textbf{TO DISCUSS}: #2}}}{}%
  	\ifstrequal{#1}{2}{\textcolor{darkcyan}{\textit{\textbf{TO DISCUSS}: #2}}}{}%
  	\ifstrequal{#1}{3}{\textcolor{darkgreen}{\textit{\textbf{TO DISCUSS}: #2}}}{}
}
\newcommand{\todo}[2]{
	\ifstrequal{#1}{0}{\textcolor{darkamber}{\textbf{\underline{TO DO}}: #2}}{}%
  	\ifstrequal{#1}{1}{\textcolor{darkmagenta}{\textbf{\underline{TO DO}}: #2}}{}%
  	\ifstrequal{#1}{2}{\textcolor{darkcyan}{\textbf{\underline{TO DO}}: #2}}{}%
  	\ifstrequal{#1}{3}{\textcolor{darkgreen}{\textbf{\underline{TO DO}}: #2}}{}
}
\newcommand{\toaddress}[2]{
	\ifstrequal{#1}{0}{\noindent\textcolor{darkamber}{$\bigstar$~\textbf{#2}}\\}{}%
  	\ifstrequal{#1}{1}{\noindent\textcolor{darkmagenta}{$\bigstar$~\textbf{#2}}\\}{}%
  	\ifstrequal{#1}{2}{\noindent\textcolor{darkcyan}{$\bigstar$~\textbf{#2}}\\}{}%
  	\ifstrequal{#1}{3}{\noindent\textcolor{darkgreen}{$\bigstar$~\textbf{#2}}\\}{}
}
\definecolor{trueblue}{rgb}{0.0, 0.45, 0.81}

\newcommand{\etal}{\textrm{et al.}\@\xspace}
\newcommand{\SSEA}{\textsc{strideSEA}\@\xspace}

\journal{}
\begin{document}
	
% Begin Front Matter
\begin{frontmatter}

% Title and Authorship
\title{\SSEA: A STRIDE-centric Security Evaluation Approach
\\[1em] \normalsize{\textbf{Author's draft for soliciting feedback - March 24, 2025}}
}

\author[sce]{Alvi Jawad\corref{cor}}
\ead{alvi.jawad@carleton.ca}
\author[sce]{Jason Jaskolka}
\ead{jason.jaskolka@carleton.ca}
\author[csit]{Ashraf Matrawy}
\ead{amatrawy@sce.carleton.ca}
\author[sce]{Mohamed Ibnkahla}
\ead{ibnkahla@sce.carleton.ca}

\cortext[cor]{Corresponding author}
\affiliation[sce]{organization={Department of Systems and Computer Engineering, Carleton University},
    addressline={1125 Colonel By Drive}, 
    city={Ottawa},
    state={Ontario},
    postcode={K1S 5B6}, 
    country={Canada}}
\affiliation[csit]{organization={School of Information Technology, Carleton University},
    addressline={1125 Colonel By Drive}, 
    city={Ottawa},
    state={Ontario},
    postcode={K1S 5B6}, 
    country={Canada}}

% Abstract	
\begin{abstract}
    Microsoft's STRIDE methodology is at the forefront of threat modeling, supporting the increasingly critical quality attribute of security in software-intensive systems. However, in a comprehensive security evaluation process, the general consensus is that the STRIDE classification is only useful for threat elicitation, isolating threat modeling from the other security evaluation activities involved in a secure software development life cycle (SDLC). We present \SSEA, a STRIDE-centric Security Evaluation Approach that integrates STRIDE as the central classification scheme into the security activities of threat modeling, attack scenario analysis, risk analysis, and countermeasure recommendation that are conducted alongside software engineering activities in secure SDLCs. The application of \SSEA is demonstrated in a real-world online immunization system case study. Using STRIDE as a single unifying thread, we bind existing security evaluation approaches in the four security activities of \SSEA to analyze (1) threats using Microsoft threat modeling tool, (2) attack scenarios using attack trees, (3) systemic risk using NASA's defect detection and prevention (DDP) technique, and (4) recommend countermeasures based on their effectiveness in reducing the most critical risks using DDP. The results include a detailed quantitative assessment of the security of the online immunization system with a clear definition of the role and advantages of integrating STRIDE in the evaluation process. Overall, the unified approach in \SSEA enables a more structured security evaluation process, allowing easier identification and recommendation of countermeasures, thus supporting the security requirements and eliciting design considerations, informing the software development life cycle of future software-based information systems.

\end{abstract}

% % Graphical abstract
% \begin{graphicalabstract}
% %\includegraphics{grabs}
% \end{graphicalabstract}

% % Research Highlights
% \begin{highlights}
% \item Research highlight 1
% \item Research highlight 2
% \end{highlights}

% Keywords

\end{frontmatter}
% End Front Matter

% Paper Body
\section{Introduction}
\label{sec:introduction}
Security evaluation has become a matter of significant concern as software-intensive systems continue to be exploited by malicious adversaries. Identifying the system security requirements and design considerations early in the SDLC is imperative for a secure software development life cycle (SDLC)~\cite{rouland_specification_2021}, where the objective is to ensure built-in security throughout the SDLC~\cite{dawson2010integrating}. Existing secure development processes such as Microsoft Secure SDLC~\cite{howard2006security}, OWASP CLASP~\cite{graham2006introduction}, and Seven Touchpoints~\cite{mcgraw2012software} propose ways to involve security methods, techniques, and tools alongside traditional SDLC activities. However, the different activities involved in such security evaluation, such as threat modeling, attack assessment, and risk analysis, burden software development teams (e.g., software developers, software engineers, and security analysts) with selecting different software tools and techniques, often making the analysis results incompatible with each other. Therefore, a unified approach that can enhance the synergy between the existing tools and techniques to minimize the effort on the development teams' part can immensely benefit the adoption of a secure SDLC.

STRIDE~\cite{shostack2008experiences} is often the starting point of many security evaluations~\cite{bygdas_evaluating_2021}, providing a systematic approach to elicit threats against a software system based on six different classes of threats.
% , namely, Spoofing, Tampering, Repudiation, Information disclosure, Denial-of-service, and Elevation of privilege. 
First publicized in 1999, the usefulness of STRIDE came to light when Microsoft performed their Windows Security Push in 2003~\cite{howard_inside_2003}, where the prime deliverable for software designers was a threat model based on STRIDE. While STRIDE can be viewed as a threat elicitation methodology, it can also be used as a threat mnemonic or threat taxonomy (called a \textit{threat classification scheme} in this work), adopted as the backbone of widely used threat modeling tools like Microsoft Threat Modeling Tool (TMT)~\cite{williams_evaluating_2015} and OWASP Threat Dragon~\cite{bygdas_evaluating_2021}. To date, STRIDE is the most prominent threat classification scheme in use by the majority of academia and the industry~\cite{hussain2014threat}. 

% In their comparison of existing threat modeling methodologies, Hussain~\etal~\cite{hussain2014threat} concluded that the majority of academia and industry use STRIDE or one of its variants.
% Even after more than two decades of its initial exposure, STRIDE is used to this day for systematic threat analysis both in academia and the industry~\cite{bygdas_evaluating_2021}.
% STRIDE has since then evolved over the years and is used to this day for systematic threat analysis both in academia and the industry~\cite{bygdas_evaluating_2021}.
% Even after more than two decades of its initial exposure, STRIDE still remains the most fine grained and comprehensive threat modeling methodology with successful integration into mature threat modeling tools~\cite{bygdas_evaluating_2021}.

% \todo{1}{Why do we need structured Security evaluation approaches? What happens to unstructured evaluation? Highly subjective? Especially, when quantitative?}

% \todo{1}{Why a STRIDE-centric approach? How is it useful?}
However, the use of the STRIDE threat classification scheme often ends at the threat modeling activity. The following security evaluation activities, such as exploitability, impact, and risk analyses and management for software systems, largely do not take the STRIDE classification into consideration~\cite{khan_stride-based_2017, macher_threat_2016, wang_systematic_2021, kavallieratos_cyber-attacks_2019, kavallieratos_managing_2020, jelacic_stride_2018, saripalli_quirc_2010, zhang_risk-level_2022}. In other cases~\cite{macher_sahara_2015, stine_cyber_2017, palanivel_risk-driven_2014}, while activities other than threat modeling implicitly inherits from STRIDE, the role and the advantages provided by the integration of the STRIDE threat classification scheme are not clearly defined. Thus, the structured view provided by STRIDE classification for threat modeling does not pervade the entire software security evaluation process, limiting the potential of STRIDE in the process. As a result, all such evaluation activities, including the final objective of such activities, i.e., the recommendation of countermeasures to support the requirements and design considerations for secure software systems, do not follow a unified approach.  

The maturity and popularity of STRIDE have inspired the development of new threat classification schemes such as LINDDUN~\cite{wuyts_linddun_2020} and extensions such as eSTRIDE~\cite{tuma_towards_2018}, STRIPED~\cite{srikumar_striped_2022}, and STRIDE$+$p~\cite{chen_modeling_2018}. In a similar vein, in this work, we assert the usefulness of the STRIDE classification scheme beyond its typical uses of threat elicitation. More explicitly, we wish to show how STRIDE has the potential to be a \emph{central classification scheme}, a classification that is used throughout the software security evaluation process to elicit threats in threat modeling, to elicit attack scenarios and their decomposition in attack scenario analysis, to elicit the risks in risk analysis, and to elicit the selection of countermeasures in countermeasure recommendation. In this work, we demonstrate such an approach by explicitly integrating the STRIDE classification into the use of existing tools and techniques to evaluate the security of a real-world eHealth system, as a software-based information system case study. 

% The evolution and success of STRIDE over the years in threat modeling have propelled us to explore whether STRIDE can be used for ``more than just threat modeling" by explicitly integrating STRIDE with other security evaluation activities. More succinctly, we wish to assess the potential of STRIDE in uniting existing techniques and tools that deal with different activities of the security evaluation process.

% STRIDE Categories Table
\begin{table*}[t]
\caption{STRIDE threat categories and related security objectives}
\label{tab:stride_categories}
\resizebox{\linewidth}{!}{%
\begin{tabular}{@{}lll@{}}
\toprule
\textbf{Threat Category} &
  \textbf{Example} &
  \textbf{Related Objective} \\ \midrule
\cellcolor[HTML]{FFFFFF}{\color[HTML]{000000} \textbf{S}poofing} &
  Impersonate a user or device within a system, e.g., by using their username or passwords &
  Authentication \\
\cellcolor[HTML]{FFFFFF}{\color[HTML]{000000} \textbf{T}ampering} &
  Maliciously modify, corrupt, or destroy data at rest or in transit, e.g., modifying data in a database &
  Integrity \\
\cellcolor[HTML]{FFFFFF}{\color[HTML]{000000} \textbf{R}epudiation} &
  Deny performing some malicious action due to lack of traceability, e.g., due to lack of auditing &
  Non-repudiation \\
\cellcolor[HTML]{FFFFFF}{\color[HTML]{000000} \textbf{I}nformation Disclosure} &
  Get unauthorized access to information at rest or in transit, e.g., to sensitive health information &
  Confidentiality \\
\cellcolor[HTML]{FFFFFF}{\color[HTML]{000000} \textbf{D}enial of Service} &
  Deny access to the services provided by a system, e.g., by overloading web servers with requests &
  Availability \\
\cellcolor[HTML]{FFFFFF}{\color[HTML]{000000} \textbf{E}levation of Privilege} &
  Obtain higher privileged access to resources than intended, e.g., gaining admin or root privileges &
  Authorization \\ \bottomrule
\end{tabular}}
\end{table*}

\noindent\textbf{Contributions:} Our main contributions are as follows:
% \JJ{REVISE to add CIP context and to align with Aims \& Scope}
\begin{itemize}
    \item \textbf{Development of a STRIDE-centric Security Evaluation Approach (\SSEA)}. \SSEA unifies various techniques and tools in each security evaluation activity around STRIDE to provide a structured approach to identify the potential threats, attack scenarios, and potential risks to a software system's security objectives. The evaluation results and identified security metrics (e.g., risk criticality) are then used to inform the selection and recommendation of countermeasures based on their effectiveness, supporting the system security requirements and eliciting design considerations to engineer future iterations of existing systems. The use of a central classification scheme (e.g., STRIDE) to streamline the entire security evaluation process is, to the best of our knowledge, a first.

    % \JJ{Further, several tools have been used in the four activities of the proposed approach. Please consider to elaborate on these tools by providing details such as the reasons that these tools are appropriate for this work.}

    \item \textbf{Application of \SSEA on a real-world eHealth case study}. We demonstrate in detail each activity of \SSEA on an online immunization system case study with validation of the results from system experts. Security evaluation of software-based health systems is scarce, and to the best of our knowledge, this is the first study that shows the application of a thorough security evaluation process for eHealth immunization systems. The evaluation results provide a security-focused perspective for better management of the SDLC of future eHealth systems. 
\end{itemize}

% \JJ{Although several similar approaches are mentioned such as eSTRIDE, it is not clear how the proposed approach is differentiated from the existing ones and what are the advantages and disadvantages over the existing ones.}
\SSEA is particularly useful for system analysts aiming to go beyond STRIDE-based threat modeling early in the SDLC to perform additional security evaluation activities systematically before deciding on mitigation schemes. Furthermore, it is useful for security tool developers as the consideration of using a central classification scheme can enhance tool synergy and integration throughout the security evaluation process.
% Additionally, it is generally useful for anyone performing a system security evaluation who needs and/or prefers a systematic process throughout.

% Our STRIDE-centric approach, with a detailed demonstration on the online immunization system, shows how STRIDE can be integrated into all activities beyond threat modeling, such as, attack scenario generation, exploitability, impact, and risk analysis, and how the recommendation of countermeasures can be based on such analyses.

The rest of the article is organized as follows. 
Section~\ref{sec:stride} elaborates on the STRIDE classification and its advantages against other potential central classification schemes.
% elaborates on the significance of STRIDE in security evaluation and compares it to other security frameworks.
Section~\ref{sec:related_work} positions our contributions in light of the related work. 
Section~\ref{sec:approach_overview} gives an overview of \SSEA. 
Section~\ref{sec:online_immunization_system_overview} details the online immunization system case study.
Section~\ref{sec:threat_modeling}--Section~\ref{sec:countermeasure_recommendation} illustrate each activity of \SSEA in detail.
Section~\ref{sec:discussion} discusses the assumptions, advantages, and limitations of \SSEA. 
Lastly, Section~\ref{sec:conclusions} concludes and briefly discusses future work.
% End Section

\section{STRIDE as a Central Classification Scheme}
\label{sec:stride}
% Begin Section

% \JJ{REVISE -- For each subsection, discuss their use in the context of CIP (related works?)}

In this section, we provide a brief overview of different categories in the STRIDE classification and detail alternative security classification schemes and their limitations.

\subsection{The STRIDE Classification}
\label{sub:stride_details}

The STRIDE classification consists of six categories of threats, as presented with examples in Table~\ref{tab:stride_categories}. STRIDE can be considered the \emph{attacker-centric view} of a system's objectives, where all STRIDE categories are aimed at causing some undesirable actions for a system. On the other hand, the related security objectives (column three) can be considered the \emph{defender-centric view} of a system, useful to identify desirable properties of a system and to know which STRIDE categories violate which security objectives. For example, maliciously modifying, corrupting, or destroying data in a system is an attacker-centric view, whereas trying to find ways to prevent such actions, i.e., to ensure the integrity of a system, is the defender-centric view. We use this duality of STRIDE to inform our proposed approach. The first three activities in \SSEA (threat modeling, attack scenario analysis, and risk analysis) use the attacker-centric view, whereas the last activity (countermeasure recommendation) is based on the defender-centric view to counter the attacker-centric view.

STRIDE has inspired the development of other threat modeling methodologies~\cite{wuyts_linddun_2020} as it is: (1) \emph{systematic}, since it classifies different cyber threats against a system based on six categories, (2) \emph{comprehensive}, as the threat analysis involves identifying threats on the security properties of authentication, integrity, non-repudiation, confidentiality, availability, and authorization by analyzing each system component and/or interaction, and (3) \emph{insightful,} as it points to existing vulnerabilities at the component level, leading to further analysis and/or selection of potential countermeasures. Recent applications of STRIDE in security evaluation include diverse fields, such as the automotive domain~\cite{macher_sahara_2015, macher_threat_2016, wang_systematic_2021}, maritime systems~\cite{kavallieratos_cyber-attacks_2019, kavallieratos_managing_2020}, healthcare systems~\cite{stine_cyber_2017}, smart grids~\cite{jelacic_stride_2018, kavallieratos_threat_2019}, cloud platforms~\cite{saripalli_quirc_2010}, and 5G networks~\cite{sattar_stride_2021}.

\subsection{Alternative Classification Schemes to STRIDE}
\label{sub:alternative_security_classification_schemes}
% \todo{1}{Discuss similar frameworks to STRIDE: CIA? STRIPED? LINDDUN? Why are we choosing STRIDE over them for unification?}
Alongside STRIDE, CIA and LINDDUN are the two primary classification schemes used in security evaluation. In contrast to the security objective-based view provided by the classic CIA triad~\cite{samonas_cia_2014}, i.e., confidentiality, integrity, and availability, STRIDE allows us to take an attacker's perspective for threat modeling. STRIDE can be considered a finer-grained categorization~\cite{howard_inside_2003} of CIA, where the objective of each STRIDE threat category can be linked to an extended CIA categorization (column three of Table~\ref{tab:stride_categories}). For example, information disclosure and tampering threats in STRIDE, respectively, are related to violations of confidentiality and integrity objectives in the CIA. However, STRIDE also includes threat categories like spoofing or repudiation, which map to authentication and non-repudiation violation, respectively, which are not included in the CIA triad.

A different aspect of threat modeling, called privacy threat modeling, includes systemic approaches, such as LINDDUN~\cite{deng_privacy_2011}, which is a mnemonic for Linkability, Identifiability, Non-repudiation, Detectability, Disclosure of information, Unawareness, and Non-compliance. While considered one of the most mature privacy threat modeling methodologies, a complete application of LINDDUN is complex and time-intensive, leading to lightweight alternatives such as LINDDUN GO~\cite{wuyts_linddun_2020}. Other privacy-related approaches have tried to gamify the threat modeling process to provide easier access to analysts. An example of this is the Elevation of Privilege (EoP) card game~\cite{shostack2014elevation}, which includes two extensions called TRIM and STRIPED that introduce additional privacy threat cards to introduce players to threat modeling~\cite{wuyts_linddun_2020}. However, all of these approaches are solely focused on modeling privacy threats, which is just one small part of all the threats that can plague a system.

An extended version of STRIDE, also known as eSTRIDE, has been suggested by Tuma~\etal~\cite{tuma_towards_2018} in an effort to reduce the inefficiencies present in STRIDE. In an empirical study~\cite{tuma_finding_2021} involving two industrial case studies, the authors claim that eSTRIDE is capable of finding twice as many high-priority threats when compared to STRIDE by focusing on the critical parts of the architecture. However, they also mention that the threat identification process is not any faster, the number of threats identified by the two methods within a given time frame are the same, and that there is a loss in the systematicity in the process of applying eSTRIDE. Another approach, called STRIDE$+$p~\cite{chen_modeling_2018}, is based on adding privacy concern of LINDDUN on top of STRIDE threats. So far, applications of eSTRIDE and STRIDE$+$p are limited, and STRIDE remains the most widely used methodology to classify threats in practice~\cite{hussain2014threat}.

% \JJ{DREAD method is also mentioned in section 2.2. Please consider to provide the necessary references. }
DREAD is an acronym that stands for Damage Potential, Reproducibility, Exploitability, Affected Users, and Discoverability. It is often used in tandem with STRIDE for a quantitative assessment of the risk presented by the threats in a STRIDE-based threat model. DREAD requires additional assumptions on each of its parameters to assess the exploitability and impact, i.e., the risk presented by those threats. However, these assumptions must be made for each identified threat, which makes the risk assessment process fairly time-consuming and highly subjective. A more objective alternative for risk assessment is Factor Analysis of Information Risk (FAIR)~\cite{freund2014_FAIR}, which allows quantitative assessment of risks in a business environment for easier decision-making. The FAIR model breaks down risks to a system into different measurable factors to allow the quantification of risk in financial terms. Some qualitative alternatives that require higher security expertise include the NIST cybersecurity framework (CSF)~\cite{shen2014nist} and ISO/IEC 27005~\cite{iso_isoiec_nodate}, among others. In this work, we use a different risk assessment methodology (refer to Section~\ref{sec:risk_analysis}) that can leverage STRIDE to provide a systemic view of the most critical risks and the most impacted security objectives while requiring fewer assumptions by the analyst.

Overall, we determined STRIDE to be the finest-grained and most flexible classification methodology in existence, which is why we use it as the central classification scheme in \SSEA.

\section{Related Work}
\label{sec:related_work}
% Begin Section
% Due to its effectiveness and mature tool support, other attempts have been made to incorporate STRIDE in different security evaluation approaches in a variety of disciplines. 

% \JJ{REVISE -- For each subsection, discuss their use in the context of CIP (related works?)}

In this section, we briefly discuss attempts from other researchers who have tried to integrate STRIDE in their proposed security evaluation approach.

\subsection{Security Evaluation Approaches using STRIDE}
\label{sub:other_security_evaluation_approaches_using_stride}
% \todo{1}{Compare this work against other SECURITY EVALUATION APPROACHES? How is it any different? What are the primary advantages?}

% Implicit inclusion of STRIDE
% \subsubsection{Implicit use of STRIDE beyond threat modeling}
\label{ssub:stride_implicitly_included}
Many studies have tried to include STRIDE in their proposed security evaluation approach that implicitly inherits different characteristics of STRIDE beyond the threat modeling activity.
% Macher et al.
Macher~\etal~\cite{macher_sahara_2015} combined the STRIDE methodology with the automotive hazard analysis and risk assessment (HARA) approach to present the SAHARA approach that analyzes the probability of occurrence and impact of security issues on safety concepts. However, the approach is limited to the automotive domain, and the analysis of impact is qualitative, not leading to the recommendation of any security controls. The SAHARA method is extended in~\cite{macher_threat_2016} to prioritize threats based on their risk level by adapting the DREAD methodology. 
% The authors use DREAD to assign a risk priority number to each threat and compare its applicability and effectiveness with the security levels defined in the traditional SAHARA method.

% Wang et al.
A more systematic approach to identifying and characterizing the risks presented to the automotive domain is presented by Wang~\etal~\cite{wang_systematic_2021}. In their risk identification module, they take a similar approach to our threat modeling activity, where system assets are identified with a data flow diagram, and threats are identified with a STRIDE threat model. The rest of the modules, called risk analysis and risk assessment, focus on various methods, such as using attack trees, the common vulnerability scoring system (CVSS), empirical equations, and parameters derived from industry standards to generate risk matrices. Their method draws heavily from existing automotive standards. For example, quantification of impact assessment parameters is done based on ISO 26262-3:2018, and the attack potential is derived from ISO/IEC 18045:2008, limiting the applicability of the method to the automotive domain. The analysis results do not lead to any recommended countermeasures.

% Stine et al.
Stine~\etal~\cite{stine_cyber_2017} used STRIDE to form a security questionnaire only in a risk scoring system for medical devices.
% Rouland et al.
Rouland~\etal~\cite{rouland_specification_2021} propose a more precise approach to formally specify, verify, and treat STRIDE threats. The approach suggests countermeasures in the form of a set of security requirements to help address detected STRIDE threats.

% Palanivel and Selvadurai
Palanivel and Selvadurai~\cite{palanivel_risk-driven_2014} used the results of STRIDE threat modeling as a measure of risk possibility, the possible threats present in each system state, in risk analysis. While their method claims to reduce the expected number of test cases due to low risk, they do not detail how the results from STRIDE threat modeling map to risk analysis. Additionally, they do not perform any attack scenario analysis or provide any recommendation for countermeasures.

While these security evaluation approaches reap some of the benefits of integrating STRIDE implicitly in their approach, the reason for inclusion is not clearly defined, and thus the benefits are unclear. In contrast, in \SSEA, we explicitly define the reasons for integrating STRIDE in each security evaluation activity and show how the integration benefits the activities and informs subsequent activities in the process.

% \JJ{REVISE -- Label steps of the approach to be referenced later in the text}
%------------------------
% Approach Steps and activities
\begin{figure*}[t!]
    \centering
    \includegraphics[width=\linewidth]{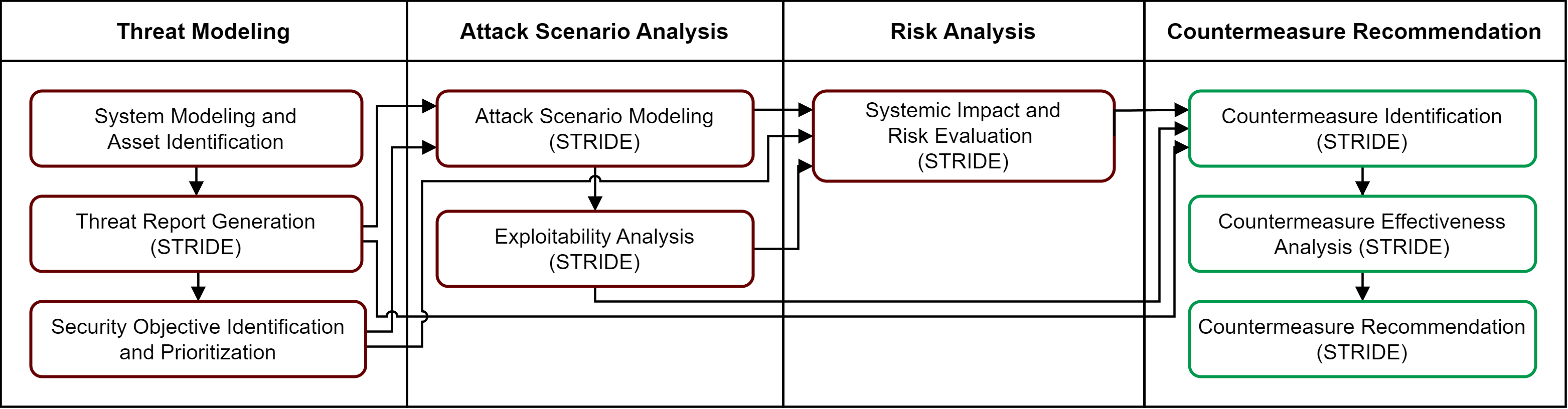}
    \caption{Steps and activities in \SSEA.}
    \label{fig:approach_steps}
\end{figure*}
%------------------------

% % Only threat modeling
% \subsubsection{STRIDE limited to threat modeling}
% \label{ssub:stride_only_threat_modeling}
% For other works that propose their own security evaluation approach, the use of STRIDE ends at the threat modeling process.
% % Saripalli and Walters
% One such work is that of Saripalli and Walters~\cite{saripalli_quirc_2010}, who presented the QUIRC framework that identifies six security objectives for cloud platforms based on STRIDE. 
% % While the threat modeling activity in QUIRC incorporates STRIDE, the rest of the activities in the framework, including risk, impact, and probability assessment, do not.
% % Zhang et al.
% % To reduce the amount of subjectivity in the risk assessment of threats,
% Zhang~\etal~\cite{zhang_risk-level_2022} suggested a modified STRIDE/DREAD model by redefining the different risk attributes and corresponding risk levels. 
% Both of these methods do not consider using STRIDE beyond the threat modeling process and are very different from the approach taken in \SSEA.

% Qualitative analyses
\subsection{Qualitative Analyses using STRIDE}
\label{ssub:qualitative_analyses}
% Kavallieratos et al.
Kavallieratos~\etal~\cite{kavallieratos_cyber-attacks_2019} use STRIDE only to categorize various attacks against systems and subsystems of cyber-enabled ships. The categorization is followed by a qualitative analysis, where they assigned low, medium, or high values to the likelihood, impact, and risk posed by an attack in a risk matrix. In a more recent work, Kavallieratos and Katsikas~\cite{kavallieratos_managing_2020} proposed a semi-quantitative risk assessment method where quantitative results are converted to qualitative data. Similar to their previous work, the threats to cyber-enabled ships were categorized based on STRIDE. However, the quantification of risk is done based on DREAD, and controls are selected without any examination of their effectiveness, which the authors admittedly leave for future work. 

% Jelacic et al.
Jelacic~\etal~\cite{jelacic_stride_2018} performed another qualitative risk analysis on a smart grid supervisory control and data acquisition (SCADA) subsystem to suggest a risk-based migration to a hybrid (community and private) cloud. The threat analysis portion is based on STRIDE, whereas the rest uses low, medium, or high values to form a risk matrix for risk analysis.

While such qualitative methods are easier to apply and allow one to better communicate the assessment results to non-technical decision makers, these methods are highly subjective and limited in the ways that they can be extended for further analysis. These limitations prompted us to opt for a quantitative approach to risk analysis in \SSEA.

Overall, while the literature is rich with threat modeling attempts based on the STRIDE classification, it is rarely used beyond the threat modeling stage in a more comprehensive security evaluation process. Additionally, most studies end at analyzing the risk to a system and only a few explore potential options for countermeasures. Even in those cases, however, the countermeasure selection process is not based on their effectiveness at reducing the risks and/or informed by the results from the analyses done up to that point. These are the exact issues that we wish to address with \SSEA. 
% End Section

\section{Overview of the Approach}
\label{sec:approach_overview}
% Begin Section
% \JJ{REVISE -- Add CIP context}

The STRIDE-centric approach proposed in this work, \SSEA, involves using a variety of techniques and tools that are suitable for security evaluation activities and can be unified around STRIDE. 
% The primary information that we need to get started is identifying the salient components in the target system of analysis and how they interact with each other and external entities during system operation. 
The approach is divided into four primary activities, as shown in Fig.~\ref{fig:approach_steps}.

% \todo{1}{ENSURE that the combination of methods does not feel ad-hoc, and that STRIDE has a definitive role to play.}
% \mynote{0}{Add a step-by-step figure to facilitate understanding. This section is IN PROGRESS.}

% \JJ{In section 4, an overview of the proposed approach is provided. However, the steps of each activity are not described. Please consider to elaborate on the steps of each activity focusing on how these steps are systematically followed to extract valid results.}

\textbf{Threat Modeling.} 
In the first \SSEA activity, we identify the potential threats to the system. We start by building a system model for analysis, where the system components and the internal and external interactions to store, process, and transmit data are modeled using a data flow diagram (DFD) ~\cite{li_data_2009} and identify the salient system assets that need protection. 
The DFD is used as input to the Microsoft TMT~\cite{williams_evaluating_2015} along with other system parameters to generate a threat report for the system classified into different STRIDE threat categories.
Lastly, we identify the security objectives of the analysis and their relative weights based on the importance of each objective. 

\textbf{Attack Scenario Analysis.} 
In the second activity, the goal is to model scenarios showing how the threats identified in the \emph{threat modeling} activity can be realized in the form of an attack. 
To this end, we present attacks against the system as attack scenarios in an attack tree, where each step leading to a successful attack is modeled as a node in the tree. The STRIDE classification maintained from the previous activity helps to develop and maintain the attack trees in a systematic manner.
% The already attacker-centric approach taken in attack trees can be facilitated by intermediate attacker goals that correspond to each threat category in STRIDE, resulting in easier identification of existing attacks that may exploit each category. 
Next, analysis of the exploitability of the attack scenarios allows us to identify the relative ease of exploiting each attack step, providing a way to prioritize attack scenarios in each STRIDE category for further analysis.
% The STRIDE-centric attack scenario identification also lends itself well to maintaining attack trees over time as system configurations change and newer attacks emerge.

% \todo{1}{Risk and Impact Analysis? Assess the possible impact and risk using DDP? Risk criticality, impacted system security objectives?}
\textbf{Risk Analysis.} 
In the third activity of \SSEA, we aim to evaluate the impact and risk posed by the identified attack scenarios on our system security objectives using NASA's Defect Detection and Prevention (DDP)~\cite{cornford_ddp-tool_2001} process. 
As input, we used the system security objectives and their relative weights, identified in the \emph{threat modeling} activity, and attack scenarios and their exploitability, identified in the \emph{attack scenario modeling} activity, to create a risk impact matrix.
The outcome of this activity is a STRIDE-centric systemic evaluation of the impact (loss of security objectives) on each security objective and the risk criticalities for each attack scenario.

% \todo{1}{Countermeasure Recommendation? Suggest countermeasures based on DDP? Countermeasure prioritization based on impacted attack scenarios?}
\textbf{Countermeasure Recommendation.} 
In the last activity, our objective is to identify and recommend suitable countermeasures for the most critical risks. 
To do so, we first identify countermeasures suitable for mitigating threats in a STRIDE category as a whole which is informed by the outcomes of all three previous activities. 
To assess the effectiveness of each countermeasure, we use the risk criticality values obtained in the \emph{risk assessment} activity along with the identified countermeasures to create a DDP countermeasures effectiveness matrix for each STRIDE category. 
The outcome is a measure of the effectiveness of each selected countermeasure or a combination of countermeasures for the most critical risks, ultimately allowing the recommendation of effective countermeasures with a STRIDE categorization.

Overall, the proposed approach uses different techniques and tools unified around STRIDE as a central classification scheme. In other words, STRIDE becomes the single thread that binds the usually distinct activities to provide a more structured security evaluation process involving: (1) threat modeling and analysis, (2) attack scenario generation and exploitability analysis, (3) risk and impact assessment, and (4) countermeasure identification and recommendation. 

% \todo{0}{\textbf{WE MAY WANT TO EMPHASIZE THIS IN THE DISCUSSION TOO:} What is important to understand here is that we may select any available techniques and tools to perform each of the activities that are suitable to analyze the target system of analysis. For example, we are not restricted to attack trees only for attack scenario generation, rather any suitable alternative could be used. It is the role of STRIDE that can be incorporated in each activity of the entire process to provide a definitive structure and other crucial benefits that we wish to draw attention to.}
% \mynote{0}{We may also wish to keep the overview more generic/open-ended, and not mention any tools or techniques in this section.}

% End Subsection
%-------------------------------------------------------------------------------------------------------------------

% \JJ{Further, several tools have been used in the four activities of the proposed approach. Please consider to elaborate on these tools by providing details such as the reasons that these tools are appropriate for this work.}

% \JJ{Several automated tools have been used in this work. It would be beneficial for the paper to provide a diagram that depicts the interactions of the tools among the steps highlighting the inputs and outputs each time that a tool is used. }

% End Section

\section{eHealth Case Study: Online Immunization System}
\label{sec:online_immunization_system_overview}
% Begin Section
% \JJ{REVISE -- Add CIP context}

\begin{figure}[ht!]
    \centering
    \includegraphics[width=0.5\linewidth]{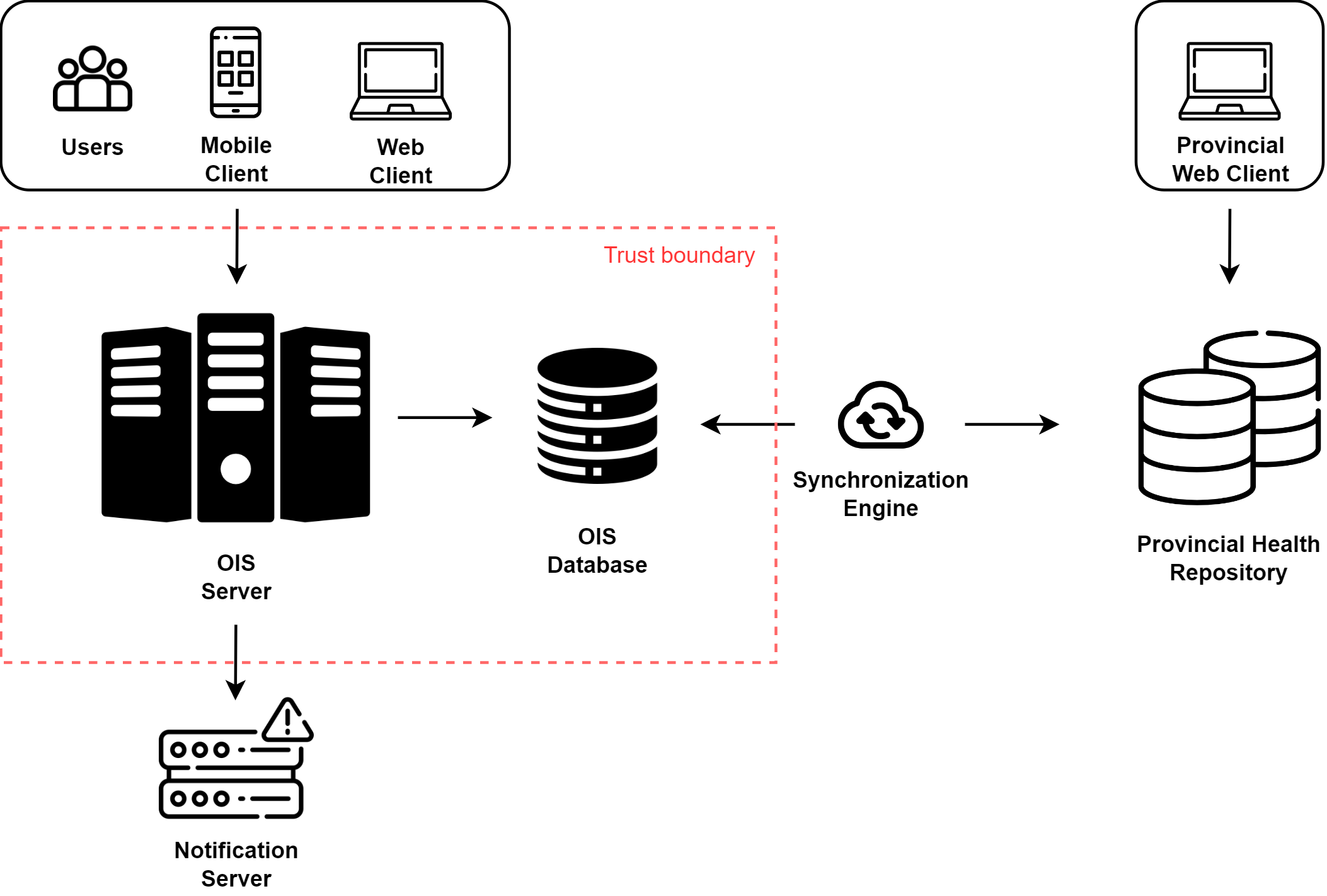}
    \caption{Architecture of the online immunization system.}
    \label{fig:ois_architecture}
\end{figure}

%-------------------------------------------------------------------------------------------------------------------
% Begin Section
% \todo{1}{Use a figure to facilitate the process. Base the figure on the CNIMNZ platform, but remove details that may cause confidentiality issues.}
In our work, we use an eHealth case study, more specifically, an online immunization system (OIS), shown in Fig.~\ref{fig:ois_architecture}, to illustrate each step in all four activities of \SSEA. The original work was done on a real-world online immunization system, specific details of which have been removed for confidentiality purposes.

%------------------------
% OIS Architecture
%------------------------`  

% \todo{1}{Describe the fundamental components and interactions of the Online Immunization System.}
The OIS consists of two primary components: the OIS server and the OIS database. Users can use the OIS mobile client or web client to interface with the OIS server to create a user account, record or update vaccination status, and access up-to-date immunization records for themselves and their families. Individual records are stored and retrieved from the OIS database. The OIS also interfaces with other regional and third-party components. Users and healthcare providers can use the provincial web client to store immunization records in the provincial health repository. A synchronization engine synchronizes individual immunization records across the OIS database and the provincial healthcare repository. The OIS can also request a third-party notification server to provide reminders to vaccinate and/or inform users of any updates to their immunization records, provided users have given their prior consent.

The components and interactions within the trust boundary are inherently trusted as they are typically physically secured and are expected to be unaffected by attacker actions. On the other hand, components outside the trust boundary are outside the control of the OIS, and interactions crossing the trust boundary are considered untrusted communication. These open up ways to exploit the system's security vulnerabilities and are the primary targets of our analysis. 
% While physical security attacks such as gaining physical access to the OIS server and subsequent vandalism are possible, it is unlikely to happen. 
Due to the highly interconnected and data-centric operation of eHealth systems, our security evaluation will focus on the cyber security aspect of the OIS, i.e., how the different type of data handled by the OIS is intercepted, manipulated, and used by attackers to attack the system, and how we can prevent it.

% End section
%-------------------------------------------------------------------------------------------------------------------
% End Section

\section{Threat Modeling}
\label{sec:threat_modeling}
% Begin Section
% \JJ{REVISE -- Add CIP context}

% \todo{1}{Discuss the fundamentals of threat modeling. The goal, inputs, and the outputs.}
Understanding the threats against a system to determine if such threats are appropriately mitigated is considered the only way to build secure software~\cite{howard_inside_2003}. Therefore, in the first activity of \SSEA, we model the threats to the OIS. The evaluation goal is \emph{to identify the salient system assets and generate a STRIDE-based threat report for such assets.}

Threat modeling is a process that can be used to identify common threats against a system's assets of interest~\cite{xiong_threat_2019}. An asset refers to anything valuable, either tangible or intangible, that needs to be protected from accidental or intentional damage or loss. The activity of threat modeling thus involves (1) identifying the salient assets within the target system of analysis that we wish to protect, (2) identifying the threats posed to a system based on STRIDE, and (3) deriving security objectives and assigning appropriate weights to them based on their role in the system. 

% \JJ{Moreover, the steps of the threat modeling are mentioned in section 6. However, the interconnections and interdependencies between the systems are not considered in the analysis. Such aspects may influence the threat modeling and risk analysis results since the propagation of the risks can be explored. Please consider to elaborate on such aspects. -- CIP interdepdendencies???}

\subsection{Asset Identification}
\label{sub:asset_identification}
% \todo{1}{Discuss the techniques and tools used. DFDs? MTMT? Emphasize the role of STRIDE!}
We adopt asset-centric threat modeling~\cite{shostack2008experiences}, where the aim is to understand the system and identify its assets that need protection. This process can be facilitated by building a system model for analysis where the system can be decomposed into constituent components to provide a clear picture of the components' interactions and operations. As the OIS primary deals with various types of data, and we wish to take a data-centric approach to assess the system security, we use \emph{data flow diagrams (DFD)}~\cite{demarco_structure_2001} in \SSEA. DFDs allow us to represent the flow of data among different system components (processes and data stores) and external entities, and visualize how data is used, processed, stored, and manipulated during operation~\cite{shostack_threat_2014}. The primary four components of a DFD are data flows, external entities, processes, and data stores~\cite{demarco_structure_2001}. 

% are shown in Fig.~\ref{fig:dfd_components}, which include,
% %------------------------
% % DFD Components
% \begin{figure}[ht!]
%     \centering
%     \includegraphics[width=\linewidth]{Figures/DFD_components.png}
%     \caption{Data flow diagram primary components~\cite{demarco_structure_2001}.}
%     \label{fig:dfd_components}
% \end{figure}
% %------------------------
% \begin{itemize}
%     \item External entity (rectangle): An outside system that communicates with the system by sending and/or receiving data. Examples include people or code outside of the system's control such as end users and external software.
%     \item Process (circle): Any function that modifies input data to produce an output. Some examples are any running code such as read/write operations and logical computations.
%     \item Data store (two parallel lines): Any form of repository or file that can preserve information for later use. Examples of data stores are databases and forms.
%     \item Data flow (arrow): The path in which data moves between external entities, processes, and data stores, typically represented as a unidirectional arrow. Bidirectional flow of data is represented with two separate unidirectional arrows.
% \end{itemize}

\begin{figure*}[t!]
    \centering
    \includegraphics[width=\linewidth]{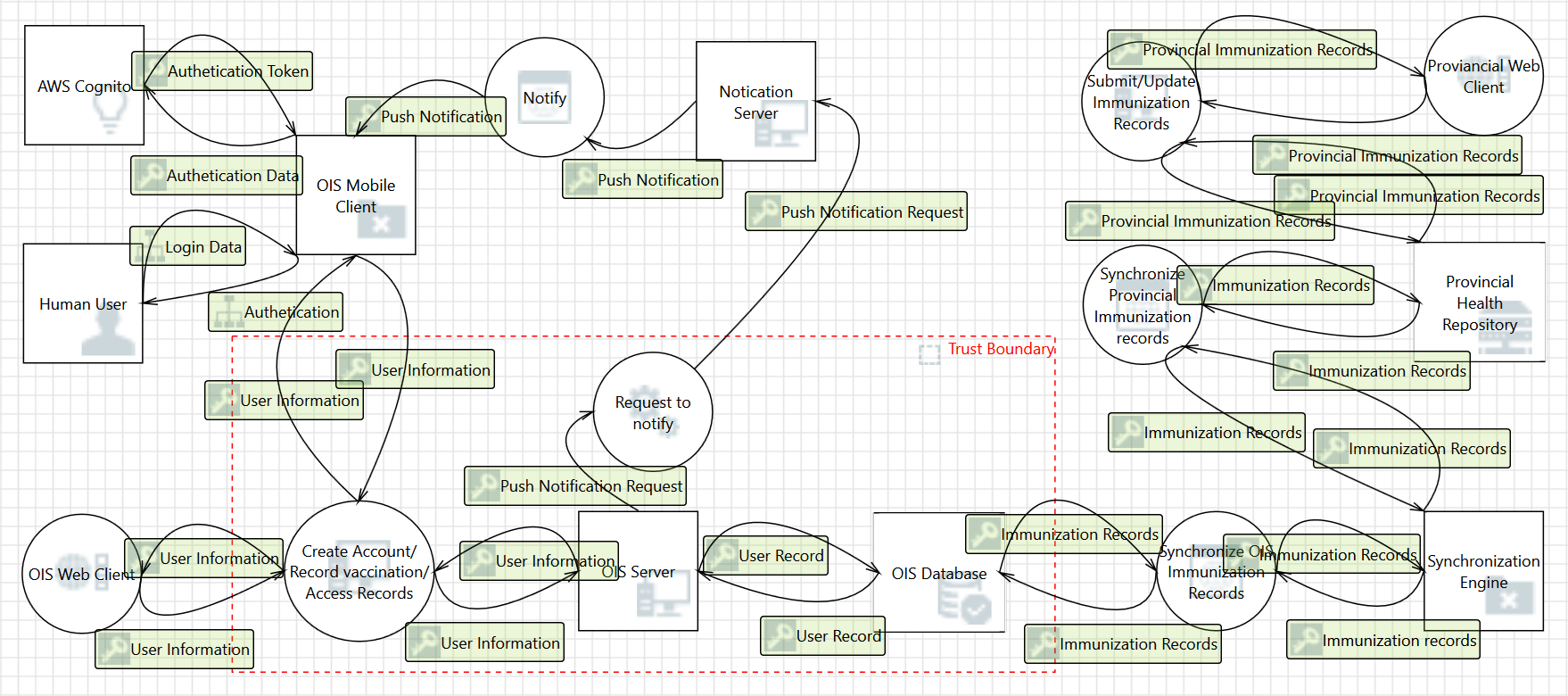}
    \caption{Data flow diagram of the online immunization system in Microsoft TMT.}
    \label{fig:ois_dfd_mtmt}
\end{figure*}

The system model of the OIS, as shown in Fig.~\ref{fig:ois_dfd_mtmt}, is modeled using the DFD notation. It highlights the different types of data exchanged (i.e., the salient assets) during the operation of the OIS, namely, 
\begin{itemize}
    \item \emph{(User) Immunization Records:} The final form of immunization information for a user to store in the provincial health repository and synchronize with the OIS.
    \item \emph{User Information:} User-provided immunization information entered using the OIS mobile or web client.
    \item \emph{User Record:} Formatted immunization records created from the \emph{User Information} to store in the OIS database.
    \item \emph{Push Notification Requests:} Notification requests sent by the OIS using a third-party server to the mobile client.
    \item \emph{Provincial Immunization Records:} Healthcare provider or user-provided immunization information entered using the provincial web client.
    \item \emph{Authentication Data/Token:} Data used for remote authentication of OIS mobile client users.
    \item \emph{Login Data:} Data used to register and log user activities.
\end{itemize}

%------------------------
% OIS DFD
%------------------------

The last three types of data are outside the trust boundary and do not interact with the OIS components directly. For example, the authentication of users using the OIS mobile client can be delegated to third-party services like AWS Cognito\footnote{https://aws.amazon.com/cognito/}, and we can ignore the external use of Login Data and Authentication Data. These two types of data are considered out of scope in our analysis. However, Provincial Immunization Records are later synchronized as Immunization Records that cross the trust boundary, and threats posed by compromise of the original Provincial Immunization Records (e.g., modification during transit or at rest) can propagate to affect the OIS operations. Therefore, our analysis in this work will focus on the first five types of data that constitute the logical assets of the system. While physical assets, such as the OIS server and database, can play important roles in other systems, due to the data-centric operation of the OIS, the cyber security aspects of the system (i.e., the logical assets) are prioritized in this work.

\subsection{Threat Report Generation}
\label{sub:threat_report_generation}
% \todo{1}{Why select MTMT among the three available tools? Justify...}
After the asset identification, the second step is to use the DFD to generate a threat model based on STRIDE. DFD is a widely used notation for system modeling for many available threat modeling tools such as Microsoft TMT~\cite{williams_evaluating_2015}, OWASP Threat Dragon~\cite{bygdas_evaluating_2021}, Mozilla SeaSponge\footnote{https://mozilla.github.io/seasponge/\#/}, and Izar PyTM\footnote{https://github.com/izar/pytm}. Among these, the first three classify threats based on STRIDE. SeaSponge was an attempt to rebuild Microsoft TMT as a cross-platform open-source tool with better graphics, which did not end up meeting the project expectations~\cite{wolf_combining_2019}. Inspired by SeaSponge, OWASP Threat dragon was developed as an open-source, multi-platform threat modeling tool, which is still in its early stages of development with no advanced threat generation mechanism available~\cite{bygdas_evaluating_2021}. Therefore, despite its steeper learning curve, we selected \emph{Microsoft TMT} as our STRIDE-based threat modeling tool as it is free and mature, provides numerous configuration options and templates, and detailed threat generation capabilities. 

The DFD of the OIS in Fig.~\ref{fig:ois_dfd_mtmt} is modeled directly in Microsoft TMT. The threat model is used to generate a threat report in \texttt{.htm} format that contains 102 unmitigated threats, one of which can be discarded as it deals with Authentication Data.

\subsection{Security Objective Identification and Prioritization}
\label{sub:security_objective_identification}

% \JJ{In section 6.3 the security objectives are described. However, it is not clearly mentioned how these objectives are identified (what documents are used, and how the security experts identified these objectives).}

% \JJ{Furthermore, it is not clear whether the security objectives are the security requirements of the system under analysis. }

After generating the threat report, we can see the number and types of threats affecting each of our identified assets. In the third step of the threat modeling activity, we derive our system security objectives. Additionally, depending on the importance of each objective for the system operation, we assign a weight from 0 to 1. The higher the number, the more important the objective is for the expected operation of the system. The derived security objectives are used later both in the second and third activities of \SSEA to perform attack scenario generation and risk analysis, respectively. For the OIS, the security objectives were identified based on available documentation and opinions from security experts, as shown in Table~\ref{tab:derived_security_objectives_weights}. We have taken a high-level asset-based approach, where protecting and/or ensuring the reliable operation of the OIS system assets is our foremost priority.
% \mynote{2}{We may simply want to say the importance of objectives, and not the weights here, especially if we do not mention specific tools in the process (DDP would be unknown at this point)}
%------------------------
% Security Objectives table
\begin{table}[ht!]
\centering
\caption{Derived security objectives for the OIS}
\label{tab:derived_security_objectives_weights}
\begin{tabular}{lll}
\hline
\textbf{No.} & \textbf{Security Objective} & \textbf{Importance} \\ \hline
1 & \cellcolor[HTML]{FFFFFF}Protecting the User Immunization Records & 1 \\
2 & Protecting the User Records & 1 \\
3 & Protecting the User Information & 0.8 \\
4 & Ensuring that the Push Notification Requests work & 0.5 \\
5 & Protecting the Provincial Immunization Records & 0.2 \\ \hline
\end{tabular}%
\end{table}
%------------------------

At the end of the first activity (Threat Modeling) of \SSEA, we are left with: \emph{(1) a model of the system components and their interactions represented as a DFD, (2) a threat report from Microsoft TMT with 101 unmitigated threats classified based on STRIDE, and (3) a list of system security objectives along with their weights of importance}.

\subsection{Discussion on Control Identification}
\label{sub:limitations_threat_modeling}
Adopting threat modeling early is useful to establish early system security requirements and examine existing controls for the identified threats~\cite{howard_inside_2003}. However, threat modeling, on its own, is not enough to recommend \emph{effective countermeasures} for a system. 

First, threat modeling tools may lead to a ``Threat Explosion'' where many threats not applicable to the system are generated. Therefore, each threat and the suggested countermeasure(s) require further investigation. For example, if no attacks exist or can be reasonably surmised as the manifestation of a threat, a countermeasure for such a threat may not be necessary. Second, at this point, we do not know the exploitability or impact, i.e., the risk presented by each threat. Suggesting countermeasures only based on the vague threat modeling results could be akin to using valuable resources on developing countermeasures for threats that are extremely unlikely to occur (very low exploitability) or have little-to-no negative effect on the system and its operations (very low impact). Therefore, we assert that countermeasures be selected after further analysis. One such approach that still retains the STRIDE-based structure is demonstrated through the rest of the activities in \SSEA.

% End Section

\section{Attack Scenario Analysis}
\label{sec:attack_scenario_analysis}
% Begin Section
% \JJ{REVISE -- Add CIP context}

% ExampleAttackTree_ImmunizationRecords
\begin{figure*}[t!]
    \centering
    \includegraphics[width=\linewidth]{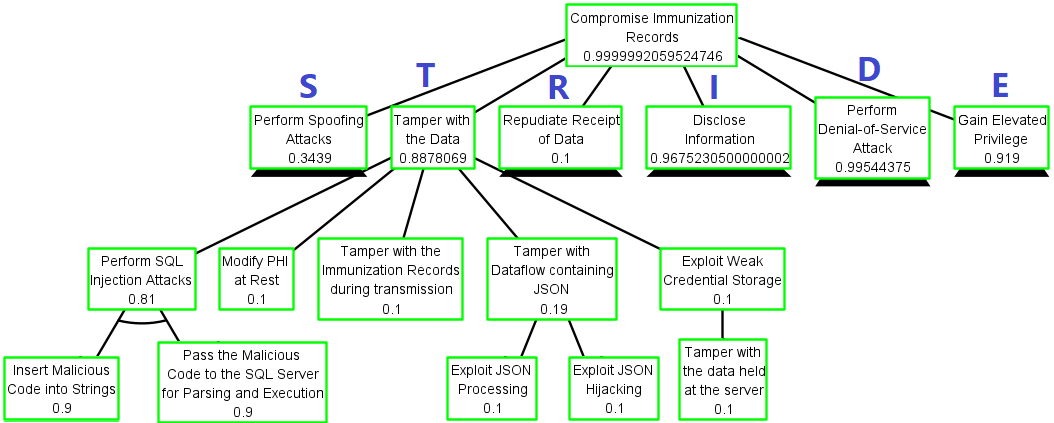}
    \caption{Attack scenarios and their exploitability for tampering with \emph{Immunization Records} in ADTool~\cite{kordy_adtool_2013}}
    \label{fig:example_attack_tree_tampering_immunization_records}
\end{figure*}
%------------------------

% \todo{1}{Discuss the fundamentals of Attack assessment. The goal, inputs, and the outputs. Attack scenarios? Attack paths? Relative exploitability?}
Once the potential threats to a system are identified in the threat modeling activity (refer to Section~\ref{sec:threat_modeling}), we need to explore corresponding attacks, i.e., ways in which such threats can manifest in the real world. In the second activity of \SSEA, we aim \emph{to perform attack scenario analysis, identifying and enumerating potential attack scenarios centered around STRIDE and assessing their exploitability for all system assets.}

\subsection{Attack Scenario Modeling}
\label{ssub:attack_scenario_modeling}
Attack scenarios are different ways in which an attacker may compromise a target asset in a system. Visualizing attack scenarios can clarify the decision-making process an attacker would go through to realize, e.g., a particular threat from the threat model of a system. They are best understood when modeled as a sequence of individual attacker actions, where reaching the attacker's goal (compromise of the target asset) is dependent on the success of all such actions. They can be modeled using threat trees~\cite{howard_inside_2003}, Bayesian networks~\cite{huang_assessing_2018}, attack trees~\cite{schneier1999attack}, attack graphs~\cite{orojloo_evaluating_2015}, among others. For our purposes, we wanted to focus on graphical representations that would enable easy visualization of different attack scenarios for people in the eHealth sector without prior security expertise. Thus, we selected \emph{attack trees} to model our attack scenarios. 

% These scenarios are usually identified once threats against the system vulnerabilities have been identified since they can drive the creation of attack scenarios. For example, attack scenarios can present the decision-making process an attacker would go through to realize, e.g., a particular threat from the threat model of a system. Identifying the many different attack scenarios against system assets is, therefore, the next step after threat modeling in a comprehensive attack analysis process. 
% Further analysis can then assess the cost, difficulty, exploitability, or skill level needed to perform each such scenario.

Attack trees present a simple visualization of attacks against a system as a tree. For modeling and simulation of attack trees, commercial software such as SecurITREE~\cite{ingoldsby2010attack} and Isograph AttackTree Software\footnote{https://www.isograph.com/software/attacktree/} are available with paid usage. Other academic tools, such as SeaMonster~\cite{meland_seamonster_nodate}, do not support quantitative analysis. We chose ADTool~\cite{kordy_adtool_2013}, a free, open-source software that enables easy creation and efficient editing of attack trees along with automated bottom-up evaluation of security-relevant measures.

An example attack tree showing tampering attack scenarios for the \emph{Immunization Records} is shown in Fig.~\ref{fig:example_attack_tree_tampering_immunization_records}. The tree is a hierarchical structure where the \emph{root} node represents the primary goal of the attacker, and its decomposition leads to \emph{branch} and \emph{leaf} nodes which depict different ways (attacker actions) of reaching the primary goal~\cite{schneier1999attack}. In complex systems, there are usually multiple different targets that are of interest to the attacker, and as such, there are multiple root nodes~\cite{kordy_adtool_2013}. In our analysis, each root node represented the compromise of a system asset (e.g., \emph{Immunization Records}), which will lead to a loss of satisfaction with the OIS security objectives identified in Table~\ref{tab:derived_security_objectives_weights}.

The role of STRIDE becomes evident once we start decomposing. Attacks on each asset can be decomposed into secondary goals based on STRIDE. For example, a secondary goal could be \emph{Tampering with the Immunization Records}, the decomposition of which can then focus on ways to perform data tampering attacks only. The decomposition of attacker actions can be represented by either conjunctions (AND nodes) or disjunctions (OR nodes). We keep decomposing each node until we reach the leaf nodes at the bottom that depict the atomic activities, i.e., the beginning of an attack attempt. Similar to the decomposed \emph{Tampering with the Immunization Records} node, we use each of the elements of STRIDE (when applicable) as an intermediate node (subgoal), the decomposition of which has been omitted due to space constraints.

The attack decomposition process can be directly guided by the STRIDE-based threat model. For example, we can look at the tampering threats for the \emph{Immunization Records} and devise ways to realize that threat from an attacker's perspective. Typically, the threat description provided by Microsoft TMT is enough to get started with the decomposition. For attack decompositions that require more information, we used community security resources such as Common Attack Pattern Enumeration and Classification (CAPEC)\footnote{https://capec.mitre.org/data/definitions/1000.html}.

% The steps at the same level of decomposition are in no particular order. 

This STRIDE-based decomposition provides a structured and modular approach to creating initial attack trees. Attack Trees can quickly become unwieldy without structure. Even for the relatively simple OIS, the \emph{Immunization Records} asset has 31 STRIDE threats associated, which keeps increasing the more complex the system is. Each of these threats will have one or more attack scenarios leading to its manifestation. If we wanted to use the realization of each of the 31 threats as a secondary goal, the resulting decomposition would easily become too complex and counteract the simple visualization provided by attack trees. With a STRIDE-based decomposition, one needs to devise attack scenarios for a smaller subgroup of identified threats (only five threats for Tampering) from the threat model.

Furthermore, maintenance of attack trees is an iterative process driven by changes or upgrades made to the system~\cite{ingoldsby2010attack}. Without any structure, such changes are not only difficult to make but also present the possibility of the tree becoming too large and complex to handle over time. The use of STRIDE-based modular subgoals can lead to efficient addition or removal of scenarios in case of changes to the system. Moreover, creating attack scenarios for a particular type of STRIDE subgoal, e.g., tampering, can lead to the development of \emph{tampering attack scenario patterns}, which can be reused when devising tampering attack scenarios for any other system asset.

\subsection{Exploitability Analysis}
\label{ssub:exploitability_analysis}
% \todo{1}{Demonstrate with figures and tables.}
Once the creation of attack scenarios is completed, it is time to analyze the scenarios. Nodes of an attack tree can be associated with values representing different properties of that attack step for further analysis. For example, the leaf nodes can be associated with the property of exploitability, difficulty, or cost of that step, and/or skill needed to perform such a step~\cite{schneier1999attack}. These properties can then be automatically calculated (e.g., using ADTool~\cite{kordy_adtool_2013}) for other nodes in the tree based on special rules of conjunction and disjunction for that property. In the second activity of \SSEA, we are interested in identifying the exploitability, i.e., how easy it is for an attacker to execute a scenario. The exploitability analysis is performed by measuring the probability of success of compromising the attacked asset. The probability of compromising the attacked asset (the root node) depends on the probability of each of the attack steps (branch nodes) and the values assigned to independent attacker actions (leaf nodes).

% Attack trees are suitable for exploitability analysis~\cite{kordy_adtool_2013}, which can be extended to risk analysis in further steps of our approach. 
% While efforts have been made to associate attack tree nodes with impact values to identify the risk of each node~\cite{edge_using_2006}, attack trees are generally not suitable for that. Not all leaf nodes can be associated with an impact value, e.g., decompositions of an AND node may not have any impacts (zero impact) associated individually, which may leaf to the nodes imposing zero risk to the system, as $risk = exploitability \times impact$.

The values representing exploitability assigned to each leaf node are based on the attack likelihood levels as described in Table~\ref{tab:attack_likelihood_levels}. The chosen values should be reasonable while satisfying the descriptions of Low, Moderate, and High. For example, the attack likelihood of Moderate represents the exploitability of attacks that are just as likely to occur as not. For a bound of $[0, 1]$, this means that a likelihood of $0.5$ is a reasonable value for attacks that have Moderate probability. On the other hand, attacks with a Low likelihood level have a low chance of occurring in normal conditions and should fall in the lower end of the $[0, 1]$ spectrum (e.g., less than $0.3$). In our experiments, the likelihood levels of Low, Moderate, and High correspond to probability measures of $0.1$, $0.5$, and $0.9$, respectively.

\begin{table}[ht!]
\centering
\caption{Attack likelihood levels and associated tree node values}
\label{tab:attack_likelihood_levels}
\begin{tabular}{@{}lll@{}}
\toprule
\textbf{\begin{tabular}[c]{@{}l@{}}Attack \\ Likelihood\end{tabular}} & \textbf{Expanded Definition} & \textbf{\begin{tabular}[c]{@{}l@{}}Node \\ value\end{tabular}} \\ \midrule
Low & \begin{tabular}[c]{@{}l@{}}Has a low chance of occurring in normal conditions. May   \\ occur in certain cases or exceptional circumstances but not \\ expected given current controls, circumstances, and recent events.\end{tabular} & 0.1 \\ \midrule
Moderate & \begin{tabular}[c]{@{}l@{}}Just as likely to occur as not. The occurrence might be   \\ difficult to control due to external influences.\end{tabular} & 0.5 \\ \midrule
High & \begin{tabular}[c]{@{}l@{}}Has a high chance of occurring considering current controls \\ and mitigations in place. Will probably occur in some, if not \\ most circumstances, and one should not be surprised if it occurred.\end{tabular} & 0.9 \\ \bottomrule
\end{tabular}%
\end{table}

In an attack tree, a node’s value is a function of its children nodes. The probability of an AND node for exploitability analysis is given by the equation~\cite{edge_using_2006},
\begin{equation}
    P_{\text{AND}} = \prod_{i=1}^n LoO_i
\end{equation}
where $i$ represents the number of child nodes and the Likelihood of Occurrence ($LoO_i$) represents the probabilities of a successful attack~\cite{edge_using_2006}. Conversely, the probability of an OR node is given by, 
\begin{equation}
    P_{\text{OR}} = 1- \prod_{i=1}^n (1 - LoO_i)
\end{equation}

For example, in Fig~\ref{fig:example_attack_tree_tampering_immunization_records}, we can say that the action \emph{Perform SQL Injection Attacks} have a higher likelihood of occurring ($0.81$) compared to \emph{Directly modify PHI at Rest} or \emph{Exploit Weak Credential Storage}, which are much less likely (both $0.1$) due to the existing security mechanisms for the OIS. Because we use STRIDE to categorize the scenarios, we can also quickly see that the likelihood of all attacks leading to successful tampering ($0.88$) is much higher compared to spoofing attacks ($0.34$) but is lower compared to information disclosure, denial-of-service, or elevation of privilege attack scenarios (all above $0.90$). The STRIDE-based decomposition gives us a quick overview of the exploitability of attack scenarios in each STRIDE category, which can be expanded to explore more detailed scenarios.
The exploitability of attack scenarios for all OIS assets is presented in Table~\ref{tab:exploitability_all_ois_assets}. In general, the likelihood of tampering, information disclosure, denial-of-service, and elevation-of-privilege attacks are higher compared to others, with denial-of-service attack scenarios being the most exploitable.

\begin{table}[ht!]
\centering
\caption{Exploitability of attack scenarios on all OIS Assets}
\label{tab:exploitability_all_ois_assets}
\begin{tabular}{@{}lllllll@{}}
\toprule
\textbf{Target Asset} & \multicolumn{6}{l}{\textbf{Exploitability Level of Attack Scenarios}}
\\ \midrule
 & \textbf{S} & \textbf{T} & \textbf{R} & \textbf{I} & \textbf{D} & \textbf{E} \\  \midrule
%  & \textbf{\parbox[t]{2mm}{{\rotatebox[origin=c]{90}{Spoofing}}}} & \textbf{\parbox[t]{2mm}{{\rotatebox[origin=c]{90}{Tampering}}}} & \textbf{\parbox[t]{2mm}{{\rotatebox[origin=c]{90}{Repudiation}}}} & \textbf{\parbox[t]{2mm}{{\rotatebox[origin=c]{90}{\begin{tabular}[c]{@{}l@{}}Information \\ Disclosure\end{tabular}}}}} & \textbf{\parbox[t]{2mm}{{\rotatebox[origin=c]{90}{\begin{tabular}[c]{@{}l@{}}Denial-of-\\ Service\end{tabular}}}}} & \textbf{\parbox[t]{2mm}{{\rotatebox[origin=c]{90}{\begin{tabular}[c]{@{}l@{}}Elevation of \\ Privilege\end{tabular}}}}} \\ \midrule
(User) Immunization Record & 0.34 & 0.88 & 0.1 & 0.96 & 0.99 & 0.91 \\
User   Records & 0.19 & 0.84 & 0.1 & 0.5 & 0.97 & N/A \\
User Information & 0.19 & 0.61 & 0.1 & 0.67 & 0.98 & 0.92 \\
Push Notification Requests & 0.19 & N/A & 0.1 & N/A & 0.95 & 0.19 \\
Provincial Immunization Records & 0.19 & 0.72 & N/A & 0.76 & 0.5 & 0.19 \\ \bottomrule
\end{tabular}%
\end{table}

% The attack likelihood levels (Low, Medium, or High) assigned to individual attacker actions are determined based on the examination of existing security documents. . 

Overall, at the end of our second \SSEA activity (Attack Scenario Analysis), we are left with: \emph{(1) a STRIDE-based decomposition of attack scenarios for all system assets modeled using attack trees and (2) the exploitability of attack scenarios in different STRIDE categories obtained through analysis of the attack trees}. It is important to understand, however, that all predictive mechanisms, including attack trees, rely on assumptions. Due to the uniqueness of the components and interactions involved in the OIS eHealth case study, it was necessary to make assumptions based on the best available information from existing security documents supported by the opinion of security experts involved in the project. For further discussion on the assumptions and limitations of attack trees, see Section~\ref{sec:discussion}.
% End Section

\section{Risk Analysis}
\label{sec:risk_analysis}
% Begin Section
% \JJ{REVISE -- Add CIP context}

Risk refers to an uncertain factor whose occurrence may result in a loss of satisfaction of a corresponding objective of a system and can be characterized as a function of the likelihood of a threat (exploitability) and the severity of its consequence(s) (impact) should it occur. Risk analysis refers to the activity of identifying and evaluating the risks faced by the system in order to mitigate such risks over the system’s lifecycle. In the third activity of \SSEA, we use the results from the threat modeling (refer to Section~\ref{sec:threat_modeling}) and attack scenario analysis (refer to Section~\ref{sec:attack_scenario_analysis}) activities \emph{to determine the STRIDE-based systemic impact and risk criticalities}.

\subsection{Defect Detection Prevention}
\label{sub:ddp}

At the heart of our risk and impact assessment lies the \emph{Defect Detection and Prevention (DDP)} mechanism~\cite{cornford_ddp-tool_2001}. DDP is a technique developed by NASA to better represent the loss of objectives through a quantitative approach for the identification, assessment, and control of identified risks. DDP's risk prioritization approach aligns closely with our goal to consider the impacts of multiple risks on multiple system objectives in a system with diverse security objectives like the OIS. It further helps optimize resource allocation towards mitigating critical risks by highlighting effective countermeasures that lead to minimal residual risks. Most importantly, DDP allows us to perform a risk assessment with high-level objectives at the beginning and use the same approach when objectives change (e.g., due to changing requirements or environment) or become more specific (e.g., more information on objectives are available) to perform a complete life-cycle risk management~\cite{cornford_ddp-tool_2001}. At later system development stages, DDP can also be integrated with other Probabilistic Risk Analysis (PRA) processes to comprise a more comprehensive risk assessment, where DDP is the ``breadth'' of the assessment and PRA provides the ``depth''~\cite{cornford_towards_2003}.

\begin{table*}[t!]
\centering
\caption{Example DDP Risk Impact Matrix for tampering attack scenarios on the OIS}
\label{tab:ddp_risk_impact_matrix_example}
\resizebox{\textwidth}{!}{%
\begin{tabular}{@{}
>{\columncolor[HTML]{FFFFFF}}l 
>{\columncolor[HTML]{FFFFFF}}l 
>{\columncolor[HTML]{FFFFFF}}l 
>{\columncolor[HTML]{FFFFFF}}l 
>{\columncolor[HTML]{FFFFFF}}l 
>{\columncolor[HTML]{FFFFFF}}l 
>{\columncolor[HTML]{FFFFFF}}l 
>{\columncolor[HTML]{FFFFFF}}l 
>{\columncolor[HTML]{FFFFFF}}l 
>{\columncolor[HTML]{FFFFFF}}l 
>{\columncolor[HTML]{FFFFFF}}l @{}}
\toprule
{\color[HTML]{333333} } & {\color[HTML]{333333} } & \multicolumn{1}{c}{\cellcolor[HTML]{FFFFFF}{\color[HTML]{333333} \textbf{Risk}}} & {\color[HTML]{333333} \begin{tabular}[c]{@{}l@{}}Perform \\ SQL\\ Injection \\ Attacks\end{tabular}} & {\color[HTML]{333333} \begin{tabular}[c]{@{}l@{}}Modify \\ PHI \\ at Rest\end{tabular}} & {\color[HTML]{333333} \begin{tabular}[c]{@{}l@{}}Tamper with\\ Immunization \\ Records during \\ transmission\end{tabular}} & {\color[HTML]{333333} \begin{tabular}[c]{@{}l@{}}Tamper with \\ Dataflow \\ containing \\ JSON\end{tabular}} & {\color[HTML]{333333} \begin{tabular}[c]{@{}l@{}}Exploit Weak \\ OIS Credential \\ Storage\end{tabular}} & {\color[HTML]{333333} \begin{tabular}[c]{@{}l@{}}Perform \\ Collision \\ Attacks\end{tabular}} & {\color[HTML]{333333} \begin{tabular}[c]{@{}l@{}}Overlap \\ Data in \\ OIS\\ Memory\end{tabular}} & \multicolumn{1}{c}{\cellcolor[HTML]{FFFFFF}{\color[HTML]{333333} \textbf{\begin{tabular}[c]{@{}c@{}}Overall\\ Impact\\ (Loss of \\ Objectives)\end{tabular}}}} \\ \midrule
\multicolumn{1}{c}{\cellcolor[HTML]{FFFFFF}{\color[HTML]{333333} \textbf{Objectives}}} & \multicolumn{1}{c}{\cellcolor[HTML]{FFFFFF}{\color[HTML]{333333} \textbf{\begin{tabular}[c]{@{}c@{}}Importance of \\      Objective\end{tabular}}}} & {\color[HTML]{333333} \textbf{\begin{tabular}[c]{@{}l@{}}Likelihood/\\      Weight\end{tabular}}} & {\color[HTML]{333333} \textbf{0.81}} & {\color[HTML]{333333} \textbf{0.1}} & {\color[HTML]{333333} \textbf{0.1}} & {\color[HTML]{333333} \textbf{0.19}} & {\color[HTML]{333333} \textbf{0.1}} & {\color[HTML]{333333} \textbf{0.5}} & {\color[HTML]{333333} \textbf{0.5}} & \multicolumn{1}{r}{\cellcolor[HTML]{FFFFFF}{\color[HTML]{333333} }} \\ \midrule
{\color[HTML]{333333} \begin{tabular}[c]{@{}l@{}}Protecting the (User)\\ Immunization Records\end{tabular}} & {\color[HTML]{333333} 1} & \multicolumn{1}{l|}{\cellcolor[HTML]{FFFFFF}{\color[HTML]{333333} \textbf{0.29}}} & {\color[HTML]{333333} 0.5} & {\color[HTML]{333333} 1} & {\color[HTML]{333333} 1} & {\color[HTML]{333333} 0.5} & {\color[HTML]{333333} 0.5} & {\color[HTML]{333333} 0} & \multicolumn{1}{l|}{\cellcolor[HTML]{FFFFFF}{\color[HTML]{333333} 0}} & {\color[HTML]{333333} \textbf{0.21}} \\
{\color[HTML]{333333} \begin{tabular}[c]{@{}l@{}}Protecting the User\\ Records\end{tabular}} & {\color[HTML]{333333} 1} & \multicolumn{1}{l|}{\cellcolor[HTML]{FFFFFF}{\color[HTML]{333333} \textbf{0.29}}} & {\color[HTML]{333333} 0.5} & {\color[HTML]{333333} 1} & {\color[HTML]{333333} 0} & {\color[HTML]{333333} 0} & {\color[HTML]{333333} 0} & {\color[HTML]{333333} 0} & \multicolumn{1}{l|}{\cellcolor[HTML]{FFFFFF}{\color[HTML]{333333} 0}} & {\color[HTML]{333333} \textbf{0.14}} \\
{\color[HTML]{333333} \begin{tabular}[c]{@{}l@{}}Protecting  the User\\ Information\end{tabular}} & {\color[HTML]{333333} 0.8} & \multicolumn{1}{l|}{\cellcolor[HTML]{FFFFFF}{\color[HTML]{333333} \textbf{0.23}}} & {\color[HTML]{333333} 0} & {\color[HTML]{333333} 0} & {\color[HTML]{333333} 0} & {\color[HTML]{333333} 0} & {\color[HTML]{333333} 0} & {\color[HTML]{333333} 0} & \multicolumn{1}{l|}{\cellcolor[HTML]{FFFFFF}{\color[HTML]{333333} 1}} & {\color[HTML]{333333} \textbf{0.11}} \\
{\color[HTML]{333333} \begin{tabular}[c]{@{}l@{}}Protecting the Provincial\\ Immunization Records\end{tabular}} & {\color[HTML]{333333} 0.5} & \multicolumn{1}{l|}{\cellcolor[HTML]{FFFFFF}{\color[HTML]{333333} \textbf{0.14}}} & {\color[HTML]{333333} 0} & {\color[HTML]{333333} 1} & {\color[HTML]{333333} 0} & {\color[HTML]{333333} 0} & {\color[HTML]{333333} 0} & {\color[HTML]{333333} 0.5} & \multicolumn{1}{l|}{\cellcolor[HTML]{FFFFFF}{\color[HTML]{333333} 0}} & {\color[HTML]{333333} \textbf{0.05}} \\
{\color[HTML]{333333} \begin{tabular}[c]{@{}l@{}}Ensuring that the Push\\ Notification Requests work\end{tabular}} & {\color[HTML]{333333} 0.2} & \multicolumn{1}{l|}{\cellcolor[HTML]{FFFFFF}{\color[HTML]{333333} \textbf{0.06}}} & {\color[HTML]{333333} 0} & {\color[HTML]{333333} 0} & {\color[HTML]{333333} 0} & {\color[HTML]{333333} 0} & {\color[HTML]{333333} 0} & {\color[HTML]{333333} 0} & \multicolumn{1}{l|}{\cellcolor[HTML]{FFFFFF}{\color[HTML]{333333} 0}} & {\color[HTML]{333333} \textbf{0.00}} \\ \midrule
{\color[HTML]{333333} } & \multicolumn{1}{c}{\cellcolor[HTML]{FFFFFF}{\color[HTML]{333333} \textbf{Risk Criticality}}} & {\color[HTML]{333333} } & {\color[HTML]{333333} \textbf{0.23}} & {\color[HTML]{333333} \textbf{0.07}} & {\color[HTML]{333333} \textbf{0.11}} & {\color[HTML]{333333} \textbf{0.03}} & {\color[HTML]{333333} \textbf{0.01}} & {\color[HTML]{333333} \textbf{0.04}} & {\color[HTML]{333333} \textbf{0.03}} & {\color[HTML]{333333} } \\ \bottomrule
\end{tabular}%
}
\end{table*}

% The primary input to DDP is our threat models and attack scenario exploitability analysis results from previous activities to obtain risk criticality and loss of objectives (impact) values. 
% Our objectives are the primary security objectives of confidentiality, integrity, and availability of assets and may also include organization-specific objectives such as patient safety. 
% In determining the system security objectives, we take a high-level asset-based approach, where protecting and/or ensuring the reliable operation of the system assets is our foremost priority, thus constituting our system security requirements (identified in Table~\ref{tab:derived_security_objectives}). 

%------------------------
% DDP Steps
\begin{figure}[ht!]
    \centering
    \includegraphics[width=0.7\linewidth]{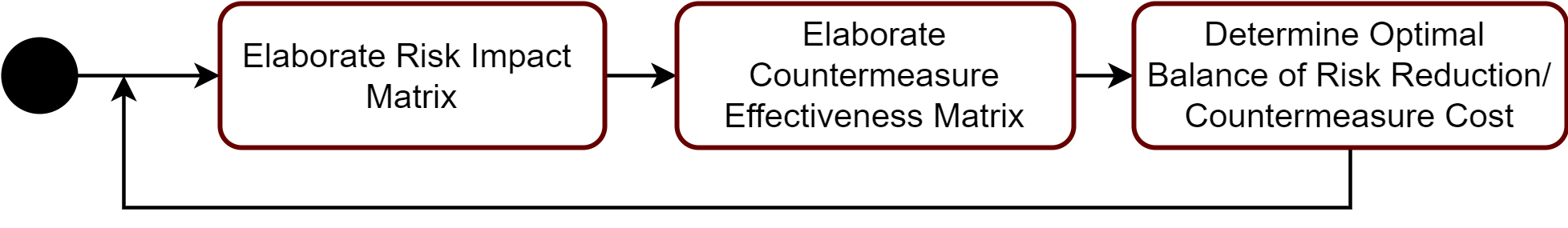}
    \caption{Steps of the DDP process.}
    \label{fig:ddp_steps}
\end{figure}
%------------------------
DDP proposes three phases for prioritizing risks and countermeasures as shown in Figure~\ref{fig:ddp_steps}: (1) elaborate a risk impact matrix, (2) elaborate a countermeasure effectiveness matrix, and (3) determine the optimal balance risk reduction/countermeasure cost. In this activity, we will focus on the first step to identify the risk criticalities and loss of objectives to determine, respectively, the most critical STRIDE risks affecting our system objectives and the most impacted objectives due to certain STRIDE attack categories. 

% The assessment results from this activity can be further used to elaborate on the effectiveness of countermeasures on the most critical risks in the last activity of \SSEA to determine the optimal balance between risk reduction and the cost of countermeasures.

% \subsection{DDP Risk Impact Matrix}
% \label{sub:ddp_risk_impact_matrix}

The \emph{DDP risk impact matrix} is a risk-consequence table where risks are prioritized by their impact on all the listed security objectives. The matrix has five components~\cite{feather_quantitative_2003}, as detailed in the example risk impact matrix in Table~\ref{tab:ddp_risk_impact_matrix_example} showing the risk presented to all OIS system assets.
\begin{enumerate}
    \item \underline{Weighted system security objectives:} The row header lists the system security objectives identified for the target system of analysis along with their relative importance for the system mission objective.
    \item \underline{Exploitable risks to the objectives:} The column header lists the risks identified for the system that may lead to the loss of the identified system objectives along with the likelihood of such risks occurring.
    \item \underline{Impact of individual risks on each objective:} Each field value in the table identifies the specific impact of each risk on each system objective and highlights objectives that are more prone to being affected by risks than others.
    \item \underline{Risk Criticality:} This is the measure of how critical each risk is when all the system objectives are concerned. The risk criticality is computed by:
    \begin{equation}
    \label{eq:4}
        Crit(r) = L(r) \times \sum_{obj} (I_{(r,obj)} \times W_{(obj)})
    \end{equation}
    where $r$ is the risk, $obj$ is the objective, $L(r)$ refers to the likelihood of occurrence for the risk, $I_{(r,obj)}$ is the impact of the risk on the system objective, and $W_{(obj)}$ is the importance of that objective relative to the other objectives.
    \item \underline{Loss of objectives:} We can measure the overall impact of all risks on each system objective by measuring the loss of an objective:
    \begin{equation}
    \label{eq:5}
        Loss(obj) = W_{(obj)} \times \sum_{r} (I_{(r,obj)}  \times L_{(r)})
    \end{equation}
    where all notations are the same as Equation~\ref{eq:4}.
\end{enumerate}

% \todo{1}{Demonstrate with figures and tables.}
In Table~\ref{tab:ddp_risk_impact_matrix_example}, we show the correspondence of the DDP risk impact matrix components with analysis of the OIS and how the previous activities of our approach inform its construction. The OIS system security objectives were identified in the first activity of our approach (see Table~\ref{tab:derived_security_objectives_weights}) along with their relative importance. The risks to the OIS were depicted by the attack scenarios that directly lead to a STRIDE attack goal (e.g., tamper with the OIS assets), and the likelihood of the risks was obtained by the exploitability analysis of the attack scenarios, both identified in the second activity (see Table~\ref{tab:exploitability_all_ois_assets}). Some attack scenarios leading to tampering with attack goals for other assets were omitted due to space constraints.

The values in the risk impact matrix fields were categorized based on the level of impact on the objective and were measured as: 
    \begin{itemize}
        \item \textbf{0}: No impact on the system objective
        \item \textbf{0.5}: Partial/low impact on the system objective; system objective may be compromised
        \item \textbf{1}: full/high impact on the system objective; system objective is compromised
    \end{itemize}
Whether a system objective can be affected by risk was determined by the existence of one or more attack scenarios in the attack trees leading to that particular risk, and the severity of the risk on the objective was determined by opinions from security experts in collaboration with OIS personnel. For further discussion on the assignment of the risk impact matrix field values, see Section~\ref{sub:assumptions_in_stridesea}.

The risk criticality, measured using Equation~\ref{eq:4}, determines how critical a risk is relative to the other identified risks based on its impact on all system objectives. For example, the most critical data tampering risk comes from \emph{performing SQL injection attacks} with a risk criticality of $0.23$ due to its high exploitability and partial impact on two of the most important security objectives of the OIS. The least critical risk ($0.01$) is presented by \emph{exploiting weak OIS credential storage}. On the other hand, Equation~\ref{eq:5} determines the extent to which a system objective is lost, given all of the risks’ impact and the likelihood of occurrence. For example, we can quickly identify that the objectives \emph{Protecting the (user) Immunization Records} and \emph{Protecting the User Records} are heavily affected, measuring at values of $0.21$ and $0.14$, respectively. Conversely, the objective \emph{Ensure that the Push Notification Requests work} is seen to be not in any kind of tampering risk at all, and therefore, there is no loss of objective.

%------------------------
% DDP graph impact (loss of objectives)
\begin{figure*}[t!]
    \centering
    \includegraphics[width=\linewidth]{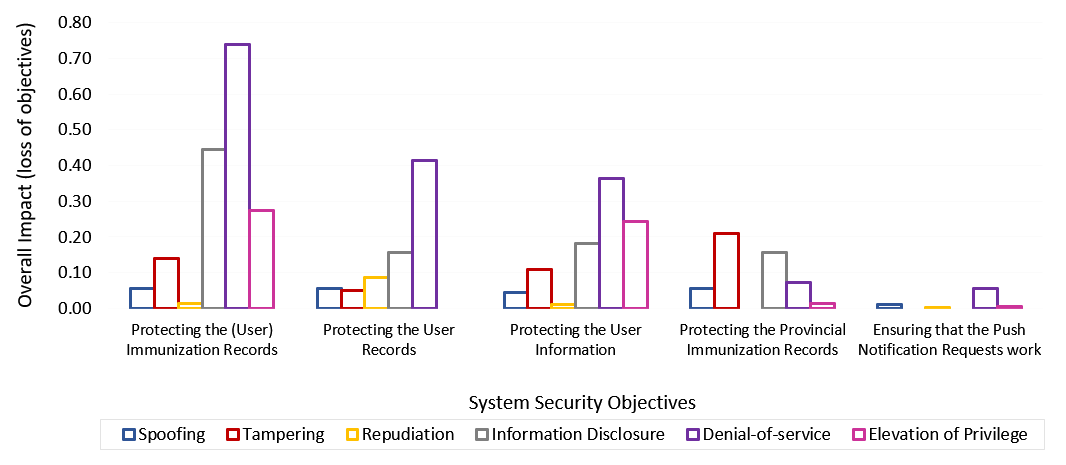}
    \caption{Overall impact (loss of objectives) on the OIS security objectives based on cumulative risks in each STRIDE category.}
    \label{fig:ddp_all_impact}
\end{figure*}
%------------------------

%------------------------
% DDP graph risk criticality
\begin{figure*}[t!]
    \centering
    \includegraphics[width=\linewidth]{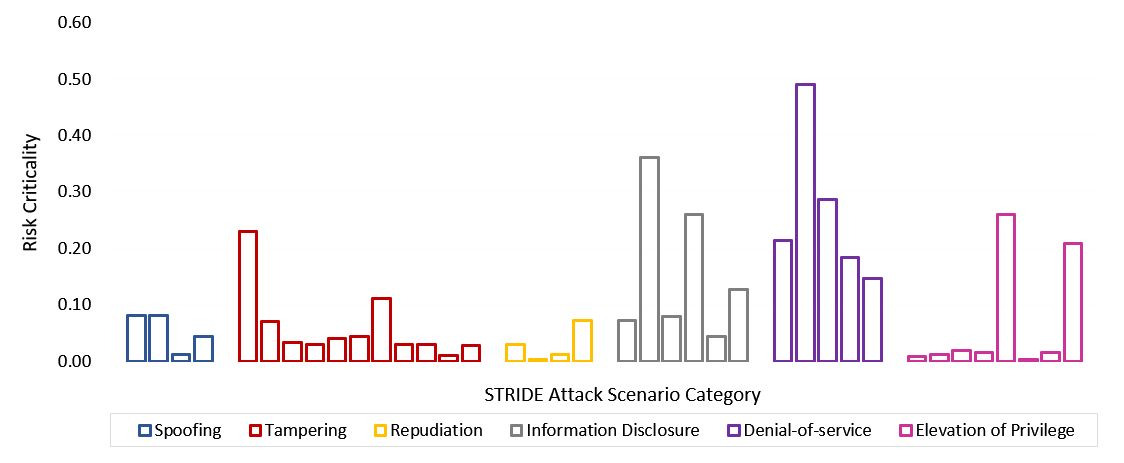}
    \caption{Risk criticality of all risks to the OIS classified based on STRIDE.}
    \label{fig:ddp_all_risk_criticality}
\end{figure*}
%------------------------

The DDP risk impact matrix highlights why basing countermeasure selection in an ad-hoc manner (e.g., immediately after threat modeling)  or on the exploitability of a risk alone (e.g., after attack scenario exploitability analysis) is not adequate; even though the risk \emph{Modify PHI at rest} has the lowest exploitability out of all ($0.1$), it is more critical ($0.07$) than many of the other risks that have higher exploitability due to its severity of impact on some of the most important OIS security objectives. In other words, taking both the exploitability and potential impact of a risk into account is of paramount importance before determining whether a countermeasure is needed and/or should be afforded for that risk. 

The use of weighted system security objectives in DDP is another point of interest that further drives the prioritization of risks. Without such weights, as long as a risk affects two objectives with the same exploitability and impact, they will exhibit equal loss of objectives. However, these two objectives may vastly differ in importance within the system, and the use of weights skews the data towards the more important objectives, drawing attention to risks that affect such objectives with a more accurate representation of the loss of objectives when considering the system operations. Similarly, the weight also influences the risk criticality values. This was one of the primary reasons for choosing DDP to do our risk analysis.

\subsection{Systemic Impact and Risk Evaluation}
\label{sub:systemic_evaluation_ddp}
The role of STRIDE becomes clear as we move towards performing a systemic evaluation with DDP. Previously, in Table~\ref{tab:ddp_risk_impact_matrix_example}, we determined the risk criticality and loss of objectives due to risks from all possible tampering attacks on all objectives. The same can be done for all STRIDE categories, which can provide a quick overview of the security posture of the target system of analysis.

Fig.~\ref{fig:ddp_all_impact} shows the overall impact (loss of objectives) on all OIS security objectives based on cumulative risks in each STRIDE category. 
Each bar in the graph shows all $38$ attack scenarios identified and categorized based on STRIDE for the OIS, the details of which are not shown here. Overall, the objective of \emph{Protecting the (User) Immunization Records} seems to be at the highest risk and contributes most to the system operations being affected. In terms of overall impact based on the STRIDE classification, denial-of-service attack scenarios comprise most of the impact, with risks from information disclosure and elevation of privilege attacks following closely after.

On the other hand, Fig.~\ref{fig:ddp_all_risk_criticality} presents a different perspective of how much of a risk the OIS is in. The risk criticality values for the different STRIDE risks reinforce our previous assertion of prioritization of countermeasures based on a combination of exploitability, impact, and importance of system security objectives. As an example, this figure incorporates all such metrics to conclude that risks from denial-of-service and information disclosure attacks are the most critical for the OIS security objectives, even though our initial assumption after the attack scenario exploitability analysis was that the exploitability of tampering, information disclosure, denial-of-service, and elevation of privilege attack scenarios was all comparable.

To summarize, at the end of our third \SSEA activity (Risk Analysis), we are left with a systemic view of:
% (1) a notion of which system security objectives are affected by which risks that are classified based on STRIDE, 
\emph{(1) Loss of system security objectives as a metric to understand the impact on system objectives, and (3) risk criticality values to understand the most critical risks in each STRIDE category that should be prioritized during mitigation.}
% End Section

\section{Countermeasure Recommendation}
\label{sec:countermeasure_recommendation}
% Begin Section
% \JJ{REVISE -- Add CIP context}

% \todo{1}{Discuss the fundamentals of countermeasure recommendation. The goal, inputs, and the outputs. Countermeasure priority? Effectiveness? Affected attack scenarios?}
The analysis done in the risk analysis activity of our approach resulted in two different systemic perspectives centered around STRIDE. 
% to view the overall impact (Fig.~\ref{fig:ddp_all_impact} and risk criticality (Fig.~\ref{fig:ddp_all_risk_criticality}) of the risks presented to the system. 
The analysis suggests two main ways to prioritize the development of countermeasures: (1) prioritize protecting the system objectives that are the most at risk, and (2) Prioritize mitigating the most critical risks. In this work, we take the second approach and perform further analysis in the last activity of \SSEA to \emph{identify suitable countermeasures based on attacks grouped by STRIDE and recommend countermeasures based on their effectiveness in reducing the risk criticality}.

\begin{table*}[t!]
\centering
\caption{Example DDP countermeasure effectiveness matrix for tampering risks to the OIS}
\label{tab:example_ddp_countermeasure_effectiveness_matrix}
\resizebox{\textwidth}{!}{%
\begin{tabular}{@{}
>{\columncolor[HTML]{FFFFFF}}l 
>{\columncolor[HTML]{FFFFFF}}l 
>{\columncolor[HTML]{FFFFFF}}l 
>{\columncolor[HTML]{FFFFFF}}l 
>{\columncolor[HTML]{FFFFFF}}l 
>{\columncolor[HTML]{FFFFFF}}l 
>{\columncolor[HTML]{FFFFFF}}l 
>{\columncolor[HTML]{FFFFFF}}l 
>{\columncolor[HTML]{FFFFFF}}l 
>{\columncolor[HTML]{FFFFFF}}l @{}}
\toprule
{\color[HTML]{000000} } & \multicolumn{1}{c}{\cellcolor[HTML]{FFFFFF}{\color[HTML]{000000} \textbf{Risk}}} & {\color[HTML]{000000} \begin{tabular}[c]{@{}l@{}}Perform \\ SQL \\ Injection \\ Attacks\end{tabular}} & {\color[HTML]{000000} \begin{tabular}[c]{@{}l@{}}Modify \\ PHI \\ at Rest\end{tabular}} & {\color[HTML]{000000} \begin{tabular}[c]{@{}l@{}}Tamper with\\ Immunization \\ Records during \\ transmission\end{tabular}} & {\color[HTML]{000000} \begin{tabular}[c]{@{}l@{}}Tamper with \\ Dataflow \\ containing \\ JSON\end{tabular}} & {\color[HTML]{000000} \begin{tabular}[c]{@{}l@{}}Exploit \\ Weak \\ Credential \\ Storage\end{tabular}} & {\color[HTML]{000000} \begin{tabular}[c]{@{}l@{}}Perform \\ Collision \\ Attacks\end{tabular}} & {\color[HTML]{000000} \begin{tabular}[c]{@{}l@{}}Overlap \\ Data in \\ Memory\end{tabular}} & {\color[HTML]{000000} \textbf{\begin{tabular}[c]{@{}l@{}}Overall \\ Effectiveness\end{tabular}}} \\ \midrule
{\color[HTML]{000000} \textbf{Selected Countermeasures}} & {\color[HTML]{000000} \textbf{Risk Criticality}} & {\color[HTML]{000000} \textbf{0.23}} & {\color[HTML]{000000} \textbf{0.07}} & {\color[HTML]{000000} \textbf{0.11}} & {\color[HTML]{000000} \textbf{0.03}} & {\color[HTML]{000000} \textbf{0.01}} & {\color[HTML]{000000} \textbf{0.04}} & {\color[HTML]{000000} \textbf{0.03}}

% \multicolumn{1}{c}{\cellcolor[HTML]{FFFFFF}{\color[HTML]{000000} \textbf{Selected Countermeasures}}} & {\color[HTML]{000000} \textbf{Risk Criticality}} & \multicolumn{1}{r}{\cellcolor[HTML]{FFFFFF}{\color[HTML]{000000} \textbf{0.23}}} & \multicolumn{1}{r}{\cellcolor[HTML]{FFFFFF}{\color[HTML]{000000} \textbf{0.07}}} & \multicolumn{1}{r}{\cellcolor[HTML]{FFFFFF}{\color[HTML]{000000} \textbf{0.11}}} & \multicolumn{1}{r}{\cellcolor[HTML]{FFFFFF}{\color[HTML]{000000} \textbf{0.03}}} & \multicolumn{1}{r}{\cellcolor[HTML]{FFFFFF}{\color[HTML]{000000} \textbf{0.01}}} & \multicolumn{1}{r}{\cellcolor[HTML]{FFFFFF}{\color[HTML]{000000} \textbf{0.04}}} & \multicolumn{1}{r}{\cellcolor[HTML]{FFFFFF}{\color[HTML]{000000} \textbf{0.03}}} & \multicolumn{1}{r}{\cellcolor[HTML]{FFFFFF}{\color[HTML]{000000} }} 

\\ \midrule
{\color[HTML]{000000} Use   cryptography} & \multicolumn{1}{l|}{\cellcolor[HTML]{FFFFFF}{\color[HTML]{000000} \textbf{}}} & {\color[HTML]{000000} 0} & {\color[HTML]{000000} 0.8} & {\color[HTML]{000000} 0.8} & {\color[HTML]{000000} 0.8} & {\color[HTML]{000000} 0.8} & {\color[HTML]{000000} 0.8} & \multicolumn{1}{l|}{\cellcolor[HTML]{FFFFFF}{\color[HTML]{000000} 0}} & {\color[HTML]{000000} \textbf{0.21}} \\
{\color[HTML]{000000} Use   appropriate access control mechanisms} & \multicolumn{1}{l|}{\cellcolor[HTML]{FFFFFF}{\color[HTML]{000000} \textbf{}}} & {\color[HTML]{000000} 0} & {\color[HTML]{000000} 0.5} & {\color[HTML]{000000} 0} & {\color[HTML]{000000} 0} & {\color[HTML]{000000} 0.5} & {\color[HTML]{000000} 0} & \multicolumn{1}{l|}{\cellcolor[HTML]{FFFFFF}{\color[HTML]{000000} 0.8}} & {\color[HTML]{000000} \textbf{0.07}} \\
{\color[HTML]{000000} Validate   and sanitize untrusted input} & \multicolumn{1}{l|}{\cellcolor[HTML]{FFFFFF}{\color[HTML]{000000} \textbf{}}} & {\color[HTML]{000000} 0.8} & {\color[HTML]{000000} 0} & {\color[HTML]{000000} 0} & {\color[HTML]{000000} 0} & {\color[HTML]{000000} 0} & {\color[HTML]{000000} 0} & \multicolumn{1}{l|}{\cellcolor[HTML]{FFFFFF}{\color[HTML]{000000} 0}} & {\color[HTML]{000000} \textbf{0.19}} \\
{\color[HTML]{000000} Use file   integrity monitoring} & \multicolumn{1}{l|}{\cellcolor[HTML]{FFFFFF}{\color[HTML]{000000} \textbf{}}} & {\color[HTML]{000000} 0} & {\color[HTML]{000000} 0.5} & {\color[HTML]{000000} 0} & {\color[HTML]{000000} 0} & {\color[HTML]{000000} 0.5} & {\color[HTML]{000000} 0} & \multicolumn{1}{l|}{\cellcolor[HTML]{FFFFFF}{\color[HTML]{000000} 0}} & {\color[HTML]{000000} \textbf{0.06}} \\ \midrule
{\color[HTML]{000000} } & {\color[HTML]{000000} \textbf{\begin{tabular}[c]{@{}l@{}}Combined \\ Risk Reduction\end{tabular}}} & {\color[HTML]{000000} \textbf{0.80}} & {\color[HTML]{000000} \textbf{0.95}} & {\color[HTML]{000000} \textbf{0.80}} & {\color[HTML]{000000} \textbf{0.80}} & {\color[HTML]{000000} \textbf{0.95}} & {\color[HTML]{000000} \textbf{0.80}} & {\color[HTML]{000000} \textbf{0.80}} & {\color[HTML]{000000} } \\ \bottomrule
\end{tabular}%
}
\end{table*}

\subsection{Countermeasure Identification}
\label{ssub:countermeasure_identification}

% \JJ{Please provide a reference for the Microsoft TMT tool that is used in section 9.1.}

% \JJ{Please provide a short discussion about the standards and guidelines that are used for the control selection in section 9.1. Please provide the systematic approach followed for the identification of the security controls.}

To recommend countermeasures, we first identify suitable countermeasures for the most critical risks. In doing so, we use the tampering risks, whose risk criticality values are already identified in the previous activity, to inform our selection. The fact that we have our risks categorized by STRIDE makes countermeasure selection a lot easier. So far, we have used the attacker-centric view (refer to Section~\ref{sub:stride_details}) provided by STRIDE since we were dealing with undesirable actions for a system such as threats, attacks, and risks. However, in case of desirable actions such as countermeasure selection, the defender-centric view is more useful. For example, we can quickly identify countermeasures that allow us to preserve the integrity of the system by countering tampering attacks. Some tools such as Microsoft TMT will even suggest generic countermeasures classified by the identified STRIDE threat category. All of these can be used as a starting point to establish a set of countermeasures with the help of existing security standards and guidelines for security control. 

By adopting this approach, we can identify a set of countermeasures for a group of threats (e.g., tampering in STRIDE) as a whole rather than trying to identify countermeasures to mitigate individual threats. This set of countermeasures can be assessed for their effectiveness (see Section~\ref{ssub:countermeasure_effectiveness_analysis}) on the existing group of threats to discard any ineffective countermeasure or identify risks not yet addressed by the current set of countermeasures. Based on the criticality of such an unaddressed risk, we can decide to either leave it unmitigated (low risk criticality) or try to identify targeted countermeasures (high risk criticality). Wider adoption of this approach can lead to the identification of different sets of countermeasures for each STRIDE category (or any other classification scheme), which can be reused or extended by further studies.

To illustrate, we identified a set of four primary countermeasures for the group of tampering attacks guided by the threat model suggestions and NIST SP 800-53~\cite{NIST_2020}:
\begin{enumerate}
    \item Use cryptography 
    \item Use appropriate access control
    \item Validate and sanitize untrusted input
    \item Use file integrity monitoring
\end{enumerate}
Using cryptographic hash function applications such as digital signatures and message authentication codes can protect the data at rest or in transit from tampering. Appropriate access control techniques can prevent unauthorized and/or unexpected modification of the memory by controlling who can access or modify sensitive files like user credentials, encryption key stores, privileges, and configuration files. None of these countermeasures, however, deal with attacker-supplied inputs, which can be mitigated by employing good programming practices to check, clean, and filter any input from an untrusted source. Additionally, file integrity monitoring techniques~\cite{peddoju_file_2020} and services, such as Tripwire~\footnote{https://www.tripwire.com/solutions}, can complement the previous techniques by monitoring any changes to sensitive files and generate an alert in case of a tampering attempt.

\subsection{Countermeasure Effectiveness Analysis}
\label{ssub:countermeasure_effectiveness_analysis}
% \todo{1}{Discuss the techniques and tools used. Nasa's DDP? Effectiveness Matrix? Spreadsheets? Emphasize the role of STRIDE!}
% As we have stressed before, the activity of countermeasure recommendation needs to be informed by the risk analysis activity to select the most effective countermeasures. 
Once selected, the countermeasures must now be assessed to determine their effectiveness in mitigating the most critical risks. DDP provides a way to assess the effectiveness of countermeasures by analyzing the capability of each countermeasure to reduce the risk criticality obtained from the DDP risk impact matrix. To that end, we use the DDP countermeasure effectiveness matrix to assess the effectiveness of the four identified countermeasures against the tampering risks to the OIS, as shown in Table~\ref{tab:example_ddp_countermeasure_effectiveness_matrix}. The DDP countermeasure effectiveness matrix has five primary components:
\begin{enumerate}
    \item \underline{Selected countermeasures:} The row header lists the countermeasures identified to deal with the presented risks.
    \item \underline{Risks and their criticality:} The column header lists the identified risks and their criticality, i.e., how critical they are for the expected system operation.
    \item \underline{Reduction in each risk due to individual countermeasures:} Each field value in the table identifies the proportion of risk reduction due to the implementation of a countermeasure.
    \item \underline{Combined risk reduction:} This is the measure of what proportion of risk is reduced when all or a subset of the selected countermeasures are taken into account. The combined risk reduction is computed by:
    \begin{equation}
    \label{eq:6}
        CRR(r) =  1 - \prod_{cm} (1 - R_{(cm,r)})
    \end{equation}
    where $r$ is the risk, $cm$ is the countermeasure, and $R_{(cm,r)}$ is the reduction in the risk due to employing the countermeasure.
    \item \underline{Overall effectiveness:} We can measure the overall effectiveness of each countermeasure in reducing all risks on the system by measuring the overall effectiveness using:
    \begin{equation}
    \label{eq:7}
        OE(cm) = \sum_r (R_{(cm,r)} \times C_{r})
    \end{equation}
    where $C_{r}$ represents the criticality of the risk and the rest of the notations are the same as Equation~\ref{eq:6}.
\end{enumerate}

In Table~\ref{tab:example_ddp_countermeasure_effectiveness_matrix}, we show how the DDP countermeasure effectiveness matrix components correspond with the previous activities in \SSEA. The countermeasure selection process is facilitated since we have a STRIDE-based decomposition throughout our process. Tampering risks, for example, aim to violate the integrity of a system, which leads to the identification of countermeasures that ensure the integrity of the OIS data and its operations. The risks and their criticality values are identified in the previous activity of risk analysis. The values in the countermeasure effectiveness matrix fields were categorized based on the level of risk reduction achieved due to employing a countermeasure and were measured as:
\begin{itemize}
    \item \textbf{0}: No reduction in risk
    \item \textbf{0.5}: Partial reduction in risk
    \item \textbf{0.8}: High reduction in risk
\end{itemize}
We use $0.8$ as the highest proportion of risk reduction based on the idea that perfect security does not exist, and even the most stringent security mechanism can fail. The values are assigned based on our optimistic view of the expected reduction for each risk. For further discussion on the assignment of the countermeasure effectiveness matrix field values, see Section~\ref{sub:assumptions_in_stridesea}.

The overall effectiveness, measured using Equation~\ref{eq:7}, indicates how effective each countermeasure is in reducing the criticality of all considered risks. This metric is affected by the risk criticality, and countermeasures that reduce high criticality risks are considered to be more effective. In case of the OIS, for example, the countermeasure \emph{use cryptography} provides a high reduction in five of the seven tampering risks considered and therefore possesses the highest effectiveness at $0.21$. On the other hand, the countermeasure \emph{validate and sanitize untrusted input} only affects one risk. However, it has the second-highest effectiveness since it provides a high reduction in the most critical tampering risk. 

Equation~\ref{eq:6} enables us to quantify the combined risk reduction, a metric that provides an understanding of how a combination of countermeasures affects risks to a system. For example, we can see that using the four selected countermeasures in tandem provides a reasonably high-risk reduction (more than $80$\%) across all the considered risks for the OIS. This metric also allows us to try a different combination of countermeasures to see how the risk reduction status of the system is affected. For example, if we were to remove the last countermeasure from the matrix, \emph{use file integrity monitoring}, it would decrease the combined risk reduction value to $0.90$ from $0.95$ for the two affected risks. However, perhaps the decreased value is still above the risk tolerance threshold for a system, and we can afford to not use that countermeasure. Removing any of the other countermeasures, however, will leave one or more of the tampering risks considered unattended.

\subsection{Countermeasure Recommendation}
\label{ssub:step3_countermeasure_recommendation}
Both the \emph{overall effectiveness} and \emph{combined risk reduction} metrics can be used to recommend countermeasures for a system. Together these metrics can help assess the benefit of implementing a set of countermeasures, the cost of which must then be taken into account to find the optimal balance between the risk reduction and countermeasure cost. The process adopted to do so involved discussion of the metrics obtained in this activity with OIS personnel and is not further detailed in this work. To mitigate tampering risks to the OIS, we recommend \emph{using cryptography} and \emph{validating and sanitizing untrusted input} as the two most important countermeasures to implement. Furthermore, \emph{using appropriate access control mechanisms} addresses the last remaining risk and puts an additional layer of security for two of the risks already addressed by the previous two countermeasures. Finally, \emph{using file integrity monitoring} is strictly optional as it has the lowest effectiveness and only marginally improves the expected risk reduction.

% Due to space constraints, we do not further our discussion on the other sets of countermeasures recommended for other STRIDE risk categories.

% The risk criticality, measured using Equation~\ref{eq:4}, determines how critical a risk is relative to the other identified risks based on its impact on all system objectives. For example, the most critical data tampering risk comes from \emph{performing SQL injection attacks} with a risk criticality of $0.23$ due to its high exploitability and partial impact on two of the most important security objectives of the OIS. The least critical risk ($0.01$) is presented by \emph{exploiting weak OIS credential storage}. On the other hand, Equation~\ref{eq:5} determines the extent to which a system objective is lost, given all of the risks’ impact and the likelihood of occurrence. For example, we can quickly identify that the objectives \emph{Protecting the (user) Immunization Records} and \emph{Protecting the User Records} are heavily affected, measuring at values of $0.21$ and $0.14$, respectively. Conversely, the objective \emph{Ensure that the Push Notification Requests work} is seen to be not in any kind of tampering risk at all, and therefore, there is no loss of objective.

Overall, at the end of the last \SSEA activity (Countermeasure Recommendation), we are left with: \emph{(1) an initial set of countermeasures selected based on the STRIDE risk under consideration (tampering), (2) the overall effectiveness of each countermeasure in reducing the risks considered, and combined risk reduction values for each risk obtained by considering single or multiple countermeasures, (4) a set of recommended countermeasures based on finding the optimal balance between the risk reduction and countermeasure cost}. The same approach can be taken to identify and recommend effective countermeasures for the other STRIDE categories as well, the details of which are not discussed here.
% End Section

\section{Discussion}
\label{sec:discussion}
% Begin Section
% \todo{1}{Discuss the results and how they are useful. Re-emphasize the use of STRIDE.}
% \mynote{2}{THIS PARAGRAPH WILL CONTAIN ASSERTIONS about how the integration of STRIDE in EACH of the security evaluation activities have streamlined the entire approach. I have ideas but I WILL write this after I see your comments on the previous two sections, which might drastically change how we present things. There will probably be at most one or two more paragraphs here.}

% \JJ{REVISE -- Add CIP context}

In this section, we briefly discuss the advantages provided by the integration of STRIDE in \SSEA, the assumptions made throughout the process, and limitations of \SSEA in its current state.

\subsection{STRIDE as the central classification scheme}
\label{sub:stride_as_central_classification_scheme}
The STRIDE classification is integrated into each activity of \SSEA, providing a structured approach to security evaluation. To that end, the tools and techniques used in this work have been purposely chosen so that they can be unified around STRIDE to provide meaningful advantages:
\begin{itemize}
    \item \underline{Threat Modeling:} For the first activity (refer to Section~\ref{sec:threat_modeling}) of \SSEA, using STRIDE is a common practice. The use of existing and mature STRIDE-based tools like Microsoft TMT, which already classifies threats based on STRIDE, allows us to automate the threat modeling process and quickly identify potential threats to a system.
    \item \underline{Attack Scenario Analysis:} In the second \SSEA activity (refer to Section~\ref{sec:attack_scenario_analysis}), potential attack scenarios are categorized for each identified system asset with a STRIDE-centric decomposition of nodes in the attack tree methodology. This approach allows us to develop patterns for attack scenarios for each STRIDE category and makes the modification, extension, and reuse of attack trees much more pragmatic over time, which are otherwise known to be cumbersome practices. The exploitability analysis enables fast identification of which STRIDE categories are most likely to be exploited by an attacker. 
    \item \underline{Risk Analysis:} The third \SSEA activity (refer to Section~\ref{sec:risk_analysis}) enables the observation of two systemic STRIDE-centric views of security: (1) the potential impact on each asset presented by risks in each STRIDE category and (2) the criticality of each risk in a STRIDE category on all system assets. The systemic views allow us to get a quick overview of the risks to our systems and determine the most alarming (impactful) risks. The two views provide different paths and analysis options for further STRIDE-based analysis of countermeasures. 
    \item \underline{Countermeasure Recommendation:} The last \SSEA activity (refer to Section~\ref{sec:countermeasure_recommendation}) is streamlined by the STRIDE categorization, as countermeasures can be systematically chosen with the use of the defender-centric view of STRIDE (see Section~\ref{sub:stride_details}). Furthermore, these countermeasures can be selected for each STRIDE category as a whole, allowing efficient initial identification of countermeasures for effectiveness analysis. Moreover, the assessment of the selected countermeasures' effectiveness is informed by steps in all of the previous STRIDE-based activities. With the effectiveness assessed, our approach can be used to recommend countermeasures against realizable threats (attacks) that pose a high risk (high exploitability and impact) to a system and not against all possible threats that are unlikely to occur or cause any meaningful impact.
\end{itemize}

The integration of different tools and techniques around a central classification scheme in \SSEA informs future security evaluation techniques and tool development to be more synergistic, enabling streamlined tool selection and analysis results to be more compatible with each other. Overall, this lessens the burden on the human side of secure software development, allowing better management and easier adoption of security practices in the SDLC.

% \mynote{0}{Please feel free to add ideas about things that you feel needs to be discussed here.}
% \todo{1}{Focus on discussing the assumptions you made and how changes to those assumptions may change the results obtained, but not the overall approach.}
% \vspace{10em}
\subsection{Assumptions in \SSEA}
\label{sub:assumptions_in_stridesea}
We made certain assumptions in \SSEA due to the techniques and tools selected for integration around STRIDE:
\begin{itemize}
    \item \underline{Attack tree node decomposition:} For quantitative analysis of attack trees in the \emph{attack scenario analysis} activity (refer to Section~\ref{sec:attack_scenario_analysis}), the decomposition of nodes has to be mutually independent, mandating the need to be careful that an attack scenario does not affect two STRIDE categories at the same time. If that is not possible, the goal of the attacker, i,e., the root node in an attack tree, can also focus on one STRIDE category at a time (e.g., tampering with asset $A$), leaving us with smaller trees with fewer attack scenarios and/or paths for each asset. All attack tree node decompositions in our work correspond to attacker actions without any overlap among nodes at the same level, thereby satisfying these assumptions.
    \item \underline{Assignment of quantitative values:} In assigning leaf node values during \emph{attack scenario analysis} (refer to Section~\ref{sec:attack_scenario_analysis}), impact values in the DDP risk impact matrix fields in \emph{risk analysis} (refer to Section~\ref{sec:risk_analysis}), and effectiveness values in the DDP countermeasure effectiveness matrix fields in \emph{countermeasure recommendation} (refer to Section~\ref{sec:countermeasure_recommendation}), we made assumptions based on the best available information and expert judgment. While the assumptions may vary based on the expertise level of an individual, in practice, such security evaluation activities will be done by a team of analysts, and a combination of experts and non-experts may reduce the potential subjectivity in taking such an approach.
    
    For example, the attack likelihood levels (Low, Medium, or High) assigned to individual attacker actions in the attack tree of Fig.~\ref{fig:example_attack_tree_tampering_immunization_records} are assumed based on the examination of existing security documents. A resulting quantitative interpretation, e.g., a likelihood of 0.88, does not mean that such an action will happen with an 88\% probability but rather how likely that attack scenario is given the chosen Low, Medium, and High values, which can then be compared to the exploitability of other attack nodes or paths. If the same analysis was done consistently with Low and High likelihoods set to 0.25 and 0.75 (or with any other reasonable value), respectively, we would have slightly different results but the same relative exploitability of attack nodes and paths. In other words, while changes in the assumptions may slightly affect the results obtained, it does not change the analysis process or activities of the proposed approach.
\end{itemize}

\subsection{Limitations of \SSEA}
\label{sub:limitations_of_stridesea}
% \todo{1}{Discuss the advantages and drawbacks of the approach. What are some of the potential pitfalls?}
The limitations of \SSEA as it stands now are as follows:
\begin{itemize}
    \item \underline{The need for security expertise}: While certain aspects of \SSEA are automated due to the inclusion of mature tools, the need for an analyst is not lost. Tool outputs can be vague, and domain or security expertise is essential in understanding and using the analysis results. Examples of this are threats in the threat report generated by Microsoft TMT, much of which requires manual observation to understand the type of compromise to the affected component(s) and whether considering such a threat for further analysis is a necessity. One possible direction of future work could explore ways to make the tool outputs more explicit and/or tailor them to specific types of systems (e.g., eHealth systems) in the form of templates, for example.
    \item \underline{Potential subjectivity}: Other challenges arise in assigning quantitative values for qualitative terms, e.g., assigning $0.1$, $0.5$, and $0.9$ to low, moderate, and high exploitability, respectively, in the attack scenario analysis activity, which is prone to subjective interpretations. Additionally, while some of the DDP matrix field values are reasonably easy to assign for an analyst (e.g., values in the countermeasure effectiveness matrix), others (e.g., values in the risk impact matrix) require collaboration with others, such as the system developers, to know the expected impact on the system objectives. Regardless, while we have tried to provide guidance on how the assignment of these values can be systematized, any other alternatives can be used as long as the analyst is consistent in their use in each activity. A future step in this regard may involve further research in how to more accurately translate the outputs from one step to the next while trying to automate as much of the approach as possible.
\end{itemize}

% \todo{1}{Discuss the challenges in performing security evaluation on eHealth systems. Highlight the real-world aspect of the examined system and perhaps some challenges related to that. Create a new subsection if necessary.}
% eHealth systems contain sensitive data, and the OIS was not an exception. We had to maintain a delicate balance while working with the OIS components to avoid accessing any confidential information that would breach privacy rules and regulations for the OIS. In doing so, our examination was limited to the different types of data that the OIS dealt with, and we never accessed any patient data directly. Throughout the process, we worked closely with OIS personnel to observe, obtain, and understand their system architecture, discuss our analysis results, and receive feedback on our recommendations, much of which was made extremely difficult due to the Covid-19 pandemic. Overall, we felt that this collaboration led to a more difficult but worthwhile security evaluation made possible by the fervent involvement of the OIS personnel, something that we would encourage future eHealth security researchers to contemplate.

% \todo{1}{Discuss some threats to validity, if time permits.}
% \vspace{15em}
% End Section

\section{Conclusions}
\label{sec:conclusions}
% Begin Section

% \JJ{REVISE -- Add CIP context}

% \todo{1}{Conclude the work and mention some ongoing/future work.}
The STRIDE classification, primarily used to categorize potential threats against a system, also has the potential to facilitate structured analysis in other security evaluation activities. In this work, we presented \SSEA, an approach where different tools and techniques are unified around STRIDE to support a more structured software security evaluation process. As a software-based health information system case study, a real-world online immunization system (OIS) is used to demonstrate the process in detail. The demonstration foregrounds the integration potential of STRIDE beyond the threat modeling activity extending to (1) attack scenario generation and their analysis using attack trees, (2) risk analysis using NASA's defect detection and prevention (DDP) technique, and (3) countermeasure recommendation based on their effectiveness in reducing the most critical risks using DDP. Our evaluation was limited to an abstract view of the different types of data that the OIS dealt with, and the evaluation results were validated by experts familiar with the OIS and its operations.

% The eHealth case study system examined in this work represents a real-world immunization system and our security evaluation results were validated by experts familiar with the OIS and its operations. In doing so, our analysis was limited to an abstract view of the different types of data that the OIS dealt with, and no patient data was accessed directly.

\SSEA highlights how the use of STRIDE as a central classification scheme allows smooth transitions from one security evaluation activity to the next, as each step is informed by one or more former steps. Such a structured approach can enable unified security evaluation throughout the SDLC, resulting in better determination of the system security requirements and design considerations to inform future considerations for engineering secure software-based information systems. Additionally, it alludes to the potential of integrating other classification schemes, such as CIA or LINDDUN, for a more streamlined security evaluation framework. In future work, we wish to explore other techniques and tools to integrate with such classification schemes and how the process can be automated for easier adoption in the secure SDLC.
% End Section

% Acknowledgement
\section*{Acknowledgement}
This work has been supported in part by the Defence Research and Development Canada’s (DRDC) Canadian Safety and Security Program (CSSP), Project \# CSSP-2018-CP-2344.

The authors would like to thank TwelveDot and Joe Samuel for providing early comments and feedback, and lending their domain expertise in validating the threat modeling phase (refer to Section~\ref{sec:threat_modeling}) of this study.

% References
\bibliographystyle{elsarticle-num} 
\bibliography{strideSEA}

\begin{thebibliography}{10}
\expandafter\ifx\csname url\endcsname\relax
  \def\url#1{\texttt{#1}}\fi
\expandafter\ifx\csname urlprefix\endcsname\relax\def\urlprefix{URL }\fi
\expandafter\ifx\csname href\endcsname\relax
  \def\href#1#2{#2} \def\path#1{#1}\fi

\bibitem{rouland_specification_2021}
Q.~Rouland, B.~Hamid, J.~Jaskolka,
  \href{https://www.sciencedirect.com/science/article/pii/S1383762121000631}{Specification,
  detection, and treatment of {STRIDE} threats for software components:
  {Modeling}, formal methods, and tool support}, Journal of Systems
  Architecture 117 (2021) 102073.
\newblock \href {https://doi.org/10.1016/j.sysarc.2021.102073}
  {\path{doi:10.1016/j.sysarc.2021.102073}}.
\newline\urlprefix\url{https://www.sciencedirect.com/science/article/pii/S1383762121000631}

\bibitem{dawson2010integrating}
M.~Dawson, D.~N. Burrell, E.~Rahim, S.~Brewster, Integrating software assurance
  into the software development life cycle (sdlc), Journal of Information
  Systems Technology and Planning 3~(6) (2010) 49--53.

\bibitem{howard2006security}
M.~Howard, S.~Lipner, The security development lifecycle, Vol.~8, Microsoft
  Press Redmond, 2006.

\bibitem{graham2006introduction}
D.~Graham, Introduction to the clasp process, Build Security In (2006).

\bibitem{mcgraw2012software}
G.~McGraw, Software security: Building security in, Datenschutz und
  Datensicherheit-DuD 36~(9) (2012) 662--665.

\bibitem{shostack2008experiences}
A.~Shostack, Experiences threat modeling at microsoft., MODSEC@ MoDELS 2008
  (2008) 35.

\bibitem{bygdas_evaluating_2021}
E.~Bygdås, L.~A. Jaatun, S.~B. Antonsen, A.~Ringen, E.~Eiring, Evaluating
  {Threat} {Modeling} {Tools}: {Microsoft} {TMT} versus {OWASP} {Threat}
  {Dragon}, in: 2021 {International} {Conference} on {Cyber} {Situational}
  {Awareness}, {Data} {Analytics} and {Assessment} ({CyberSA}), 2021, pp. 1--7.
\newblock \href {https://doi.org/10.1109/CyberSA52016.2021.9478215}
  {\path{doi:10.1109/CyberSA52016.2021.9478215}}.

\bibitem{howard_inside_2003}
M.~Howard, S.~Lipner,
  \href{https://ieeexplore.ieee.org/document/1176996/}{Inside the {Windows}
  security push}, IEEE Security \& Privacy 1~(1) (2003) 57--61.
\newblock \href {https://doi.org/10.1109/MSECP.2003.1176996}
  {\path{doi:10.1109/MSECP.2003.1176996}}.
\newline\urlprefix\url{https://ieeexplore.ieee.org/document/1176996/}

\bibitem{williams_evaluating_2015}
I.~Williams, X.~Yuan, \href{https://doi.org/10.1145/2885990.2885999}{Evaluating
  the effectiveness of {Microsoft} threat modeling tool}, in: Proceedings of
  the 2015 {Information} {Security} {Curriculum} {Development} {Conference},
  {InfoSec} '15, Association for Computing Machinery, New York, NY, USA, 2015,
  pp. 1--6.
\newblock \href {https://doi.org/10.1145/2885990.2885999}
  {\path{doi:10.1145/2885990.2885999}}.
\newline\urlprefix\url{https://doi.org/10.1145/2885990.2885999}

\bibitem{hussain2014threat}
S.~Hussain, A.~Kamal, S.~Ahmad, G.~Rasool, S.~Iqbal, Threat modelling
  methodologies: a survey, Sci. Int.(Lahore) 26~(4) (2014) 1607--1609.

\bibitem{khan_stride-based_2017}
R.~Khan, K.~McLaughlin, D.~Laverty, S.~Sezer, {STRIDE}-based threat modeling
  for cyber-physical systems, in: 2017 {IEEE} {PES} {Innovative} {Smart} {Grid}
  {Technologies} {Conference} {Europe} ({ISGT}-{Europe}), 2017, pp. 1--6.
\newblock \href {https://doi.org/10.1109/ISGTEurope.2017.8260283}
  {\path{doi:10.1109/ISGTEurope.2017.8260283}}.

\bibitem{macher_threat_2016}
G.~Macher, E.~Armengaud, E.~Brenner, C.~Kreiner,
  \href{https://www.sciencedirect.com/science/article/pii/S1877050916303015}{Threat
  and {Risk} {Assessment} {Methodologies} in the {Automotive} {Domain}},
  Procedia Computer Science 83 (2016) 1288--1294.
\newblock \href {https://doi.org/10.1016/j.procs.2016.04.268}
  {\path{doi:10.1016/j.procs.2016.04.268}}.
\newline\urlprefix\url{https://www.sciencedirect.com/science/article/pii/S1877050916303015}

\bibitem{wang_systematic_2021}
Y.~Wang, Y.~Wang, H.~Qin, H.~Ji, Y.~Zhang, J.~Wang,
  \href{https://doi.org/10.1007/s42154-021-00140-6}{A {Systematic} {Risk}
  {Assessment} {Framework} of {Automotive} {Cybersecurity}}, Automotive
  Innovation 4~(3) (2021) 253--261.
\newblock \href {https://doi.org/10.1007/s42154-021-00140-6}
  {\path{doi:10.1007/s42154-021-00140-6}}.
\newline\urlprefix\url{https://doi.org/10.1007/s42154-021-00140-6}

\bibitem{kavallieratos_cyber-attacks_2019}
G.~Kavallieratos, S.~Katsikas, V.~Gkioulos, Cyber-{Attacks} {Against} the
  {Autonomous} {Ship}, in: S.~K. Katsikas, F.~Cuppens, N.~Cuppens,
  C.~Lambrinoudakis, A.~Antón, S.~Gritzalis, J.~Mylopoulos, C.~Kalloniatis
  (Eds.), Computer {Security}, Lecture {Notes} in {Computer} {Science},
  Springer International Publishing, Cham, 2019, pp. 20--36.
\newblock \href {https://doi.org/10.1007/978-3-030-12786-2_2}
  {\path{doi:10.1007/978-3-030-12786-2_2}}.

\bibitem{kavallieratos_managing_2020}
G.~Kavallieratos, S.~Katsikas,
  \href{https://www.mdpi.com/2077-1312/8/10/768}{Managing {Cyber} {Security}
  {Risks} of the {Cyber}-{Enabled} {Ship}}, Journal of Marine Science and
  Engineering 8~(10) (2020) 768, number: 10 Publisher: Multidisciplinary
  Digital Publishing Institute.
\newblock \href {https://doi.org/10.3390/jmse8100768}
  {\path{doi:10.3390/jmse8100768}}.
\newline\urlprefix\url{https://www.mdpi.com/2077-1312/8/10/768}

\bibitem{jelacic_stride_2018}
B.~Jelacic, D.~Rosic, I.~Lendak, M.~Stanojevic, S.~Stoja, {STRIDE} to a
  {Secure} {Smart} {Grid} in a {Hybrid} {Cloud}, in: S.~K. Katsikas,
  F.~Cuppens, N.~Cuppens, C.~Lambrinoudakis, C.~Kalloniatis, J.~Mylopoulos,
  A.~Antón, S.~Gritzalis (Eds.), Computer {Security}, Lecture {Notes} in
  {Computer} {Science}, Springer International Publishing, Cham, 2018, pp.
  77--90.
\newblock \href {https://doi.org/10.1007/978-3-319-72817-9_6}
  {\path{doi:10.1007/978-3-319-72817-9_6}}.

\bibitem{saripalli_quirc_2010}
P.~Saripalli, B.~Walters,
  \href{http://ieeexplore.ieee.org/document/5557981/}{{QUIRC}: {A}
  {Quantitative} {Impact} and {Risk} {Assessment} {Framework} for {Cloud}
  {Security}}, in: 2010 {IEEE} 3rd {International} {Conference} on {Cloud}
  {Computing}, IEEE, Miami, FL, USA, 2010, pp. 280--288.
\newblock \href {https://doi.org/10.1109/CLOUD.2010.22}
  {\path{doi:10.1109/CLOUD.2010.22}}.
\newline\urlprefix\url{http://ieeexplore.ieee.org/document/5557981/}

\bibitem{zhang_risk-level_2022}
L.~Zhang, A.~Taal, R.~Cushing, C.~de~Laat, P.~Grosso, A risk-level assessment
  system based on the {STRIDE}/{DREAD} model for digital data marketplaces,
  International Journal of Information Security 21~(3) (2022) 509--525.
\newblock \href {https://doi.org/10.1007/s10207-021-00566-3}
  {\path{doi:10.1007/s10207-021-00566-3}}.

\bibitem{macher_sahara_2015}
G.~Macher, H.~Sporer, R.~Berlach, E.~Armengaud, C.~Kreiner, {SAHARA}: {A}
  security-aware hazard and risk analysis method, in: 2015 {Design},
  {Automation} \& {Test} in {Europe} {Conference} \& {Exhibition} ({DATE}),
  2015, pp. 621--624, iSSN: 1558-1101.
\newblock \href {https://doi.org/10.7873/DATE.2015.0622}
  {\path{doi:10.7873/DATE.2015.0622}}.

\bibitem{stine_cyber_2017}
I.~Stine, M.~Rice, S.~Dunlap, J.~Pecarina,
  \href{https://www.sciencedirect.com/science/article/pii/S187454821730063X}{A
  cyber risk scoring system for medical devices}, International Journal of
  Critical Infrastructure Protection 19 (2017) 32--46.
\newblock \href {https://doi.org/10.1016/j.ijcip.2017.04.001}
  {\path{doi:10.1016/j.ijcip.2017.04.001}}.
\newline\urlprefix\url{https://www.sciencedirect.com/science/article/pii/S187454821730063X}

\bibitem{palanivel_risk-driven_2014}
M.~Palanivel, K.~Selvadurai,
  \href{https://doi.org/10.1186/2193-1801-3-754}{Risk-driven security testing
  using risk analysis with threat modeling approach}, SpringerPlus 3~(1) (2014)
  754.
\newblock \href {https://doi.org/10.1186/2193-1801-3-754}
  {\path{doi:10.1186/2193-1801-3-754}}.
\newline\urlprefix\url{https://doi.org/10.1186/2193-1801-3-754}

\bibitem{wuyts_linddun_2020}
K.~Wuyts, L.~Sion, W.~Joosen, {LINDDUN} {GO}: {A} {Lightweight} {Approach} to
  {Privacy} {Threat} {Modeling}, in: 2020 {IEEE} {European} {Symposium} on
  {Security} and {Privacy} {Workshops} ({EuroS}\&{PW}), 2020, pp. 302--309.
\newblock \href {https://doi.org/10.1109/EuroSPW51379.2020.00047}
  {\path{doi:10.1109/EuroSPW51379.2020.00047}}.

\bibitem{tuma_towards_2018}
K.~Tuma, R.~Scandariato, M.~Widman, C.~Sandberg, Towards {Security} {Threats}
  that {Matter}, in: S.~K. Katsikas, F.~Cuppens, N.~Cuppens, C.~Lambrinoudakis,
  C.~Kalloniatis, J.~Mylopoulos, A.~Antón, S.~Gritzalis (Eds.), Computer
  {Security}, Lecture {Notes} in {Computer} {Science}, Springer International
  Publishing, Cham, 2018, pp. 47--62.
\newblock \href {https://doi.org/10.1007/978-3-319-72817-9_4}
  {\path{doi:10.1007/978-3-319-72817-9_4}}.

\bibitem{srikumar_striped_2022}
K.~Srikumar, K.~Kashish, K.~Eggers, N.~E. Díaz~Ferreyra, J.~Koch,
  T.~Schüppstuhl, R.~Scandariato,
  \href{https://doi.org/10.1145/3538969.3538970}{{STRIPED}: {A} {Threat}
  {Analysis} {Method} for {IoT} {Systems}}, in: Proceedings of the 17th
  {International} {Conference} on {Availability}, {Reliability} and {Security},
  {ARES} '22, Association for Computing Machinery, New York, NY, USA, 2022, pp.
  1--6.
\newblock \href {https://doi.org/10.1145/3538969.3538970}
  {\path{doi:10.1145/3538969.3538970}}.
\newline\urlprefix\url{https://doi.org/10.1145/3538969.3538970}

\bibitem{chen_modeling_2018}
Y.-T. Chen, Modeling {Information} {Security} {Threats} for {Smart} {Grid}
  {Applications} by {Using} {Software} {Engineering} and {Risk} {Management},
  in: 2018 {IEEE} {International} {Conference} on {Smart} {Energy} {Grid}
  {Engineering} ({SEGE}), 2018, pp. 128--132, iSSN: 2575-2693.
\newblock \href {https://doi.org/10.1109/SEGE.2018.8499431}
  {\path{doi:10.1109/SEGE.2018.8499431}}.

\bibitem{kavallieratos_threat_2019}
G.~Kavallieratos, V.~Gkioulos, S.~K. Katsikas, Threat {Analysis} in {Dynamic}
  {Environments}: {The} {Case} of the {Smart} {Home}, in: 2019 15th
  {International} {Conference} on {Distributed} {Computing} in {Sensor}
  {Systems} ({DCOSS}), 2019, pp. 234--240, iSSN: 2325-2944.
\newblock \href {https://doi.org/10.1109/DCOSS.2019.00060}
  {\path{doi:10.1109/DCOSS.2019.00060}}.

\bibitem{sattar_stride_2021}
D.~Sattar, A.~H. Vasoukolaei, P.~Crysdale, A.~Matrawy, A {STRIDE} {Threat}
  {Model} for {5G} {Core} {Slicing}, in: 2021 {IEEE} 4th {5G} {World} {Forum}
  ({5GWF}), 2021, pp. 247--252.
\newblock \href {https://doi.org/10.1109/5GWF52925.2021.00050}
  {\path{doi:10.1109/5GWF52925.2021.00050}}.

\bibitem{samonas_cia_2014}
S.~Samonas, D.~Coss, The cia strikes back: Redefining confidentiality,
  integrity and availability in security., Journal of Information System
  Security 10~(3) (2014).

\bibitem{deng_privacy_2011}
M.~Deng, K.~Wuyts, R.~Scandariato, B.~Preneel, W.~Joosen,
  \href{https://doi.org/10.1007/s00766-010-0115-7}{A privacy threat analysis
  framework: supporting the elicitation and fulfillment of privacy
  requirements}, Requirements Engineering 16~(1) (2011) 3--32.
\newblock \href {https://doi.org/10.1007/s00766-010-0115-7}
  {\path{doi:10.1007/s00766-010-0115-7}}.
\newline\urlprefix\url{https://doi.org/10.1007/s00766-010-0115-7}

\bibitem{shostack2014elevation}
A.~Shostack, Elevation of privilege: Drawing developers into threat modeling,
  in: 2014 USENIX Summit on Gaming, Games, and Gamification in Security
  Education (3GSE 14), 2014.

\bibitem{tuma_finding_2021}
K.~Tuma, C.~Sandberg, U.~Thorsson, M.~Widman, T.~Herpel, R.~Scandariato,
  \href{https://www.sciencedirect.com/science/article/pii/S016412122100100X}{Finding
  security threats that matter: {Two} industrial case studies}, Journal of
  Systems and Software 179 (2021) 111003.
\newblock \href {https://doi.org/10.1016/j.jss.2021.111003}
  {\path{doi:10.1016/j.jss.2021.111003}}.
\newline\urlprefix\url{https://www.sciencedirect.com/science/article/pii/S016412122100100X}

\bibitem{freund2014_FAIR}
J.~Freund, J.~Jones, Measuring and managing information risk: a FAIR approach,
  Butterworth-Heinemann, 2014.

\bibitem{shen2014nist}
L.~Shen, The nist cybersecurity framework: Overview and potential impacts,
  Scitech Lawyer 10~(4) (2014) 16.

\bibitem{iso_isoiec_nodate}
{ISO}, \href{https://www.iso.org/standard/80585.html}{{ISO}/{IEC} 27005:2022},
  accessed: 2023-03-24.
\newline\urlprefix\url{https://www.iso.org/standard/80585.html}

\bibitem{li_data_2009}
Q.~Li, Y.-L. Chen, \href{https://doi.org/10.1007/978-3-540-89556-5_4}{Data
  {Flow} {Diagram}}, in: Q.~Li, Y.-L. Chen (Eds.), Modeling and {Analysis} of
  {Enterprise} and {Information} {Systems}: {From} {Requirements} to
  {Realization}, Springer, Berlin, Heidelberg, 2009, pp. 85--97.
\newblock \href {https://doi.org/10.1007/978-3-540-89556-5_4}
  {\path{doi:10.1007/978-3-540-89556-5_4}}.
\newline\urlprefix\url{https://doi.org/10.1007/978-3-540-89556-5_4}

\bibitem{cornford_ddp-tool_2001}
S.~Cornford, M.~Feather, K.~Hicks, {DDP}-a tool for life-cycle risk management,
  in: 2001 {IEEE} {Aerospace} {Conference} {Proceedings} ({Cat}.
  {No}.{01TH8542}), Vol.~1, 2001, pp. 1/441--1/451 vol.1.
\newblock \href {https://doi.org/10.1109/AERO.2001.931736}
  {\path{doi:10.1109/AERO.2001.931736}}.

\bibitem{xiong_threat_2019}
W.~Xiong, R.~Lagerström,
  \href{https://www.sciencedirect.com/science/article/pii/S0167404818307478}{Threat
  modeling – {A} systematic literature review}, Computers \& Security 84
  (2019) 53--69.
\newblock \href {https://doi.org/10.1016/j.cose.2019.03.010}
  {\path{doi:10.1016/j.cose.2019.03.010}}.
\newline\urlprefix\url{https://www.sciencedirect.com/science/article/pii/S0167404818307478}

\bibitem{demarco_structure_2001}
T.~DeMarco, \href{https://doi.org/10.1007/978-3-642-48354-7_9}{Structure
  {Analysis} and {System} {Specification}}, in: M.~Broy, E.~Denert (Eds.),
  Pioneers and {Their} {Contributions} to {Software} {Engineering}: sd\&m
  {Conference} on {Software} {Pioneers}, {Bonn}, {June} 28/29, 2001, {Original}
  {Historic} {Contributions}, Springer, Berlin, Heidelberg, 2001, pp. 255--288.
\newblock \href {https://doi.org/10.1007/978-3-642-48354-7_9}
  {\path{doi:10.1007/978-3-642-48354-7_9}}.
\newline\urlprefix\url{https://doi.org/10.1007/978-3-642-48354-7_9}

\bibitem{shostack_threat_2014}
A.~Shostack, Threat {Modeling}: {Designing} for {Security}, John Wiley \& Sons,
  2014, google-Books-ID: YiHcAgAAQBAJ.

\bibitem{wolf_combining_2019}
M.~Wolf, Combining safety and security threat modeling to improve automotive
  penetration testing, Master's thesis, Universit{\"a}t Ulm (2019).

\bibitem{kordy_adtool_2013}
B.~Kordy, P.~Kordy, S.~Mauw, P.~Schweitzer, {ADTool}: {Security} {Analysis}
  with {Attack}–{Defense} {Trees}, in: K.~Joshi, M.~Siegle, M.~Stoelinga,
  P.~R. D’Argenio (Eds.), Quantitative {Evaluation} of {Systems}, Lecture
  {Notes} in {Computer} {Science}, Springer, Berlin, Heidelberg, 2013, pp.
  173--176.
\newblock \href {https://doi.org/10.1007/978-3-642-40196-1_15}
  {\path{doi:10.1007/978-3-642-40196-1_15}}.

\bibitem{huang_assessing_2018}
K.~Huang, C.~Zhou, Y.-C. Tian, S.~Yang, Y.~Qin, Assessing the {Physical}
  {Impact} of {Cyberattacks} on {Industrial} {Cyber}-{Physical} {Systems}, IEEE
  Transactions on Industrial Electronics 65~(10) (2018) 8153--8162.
\newblock \href {https://doi.org/10.1109/TIE.2018.2798605}
  {\path{doi:10.1109/TIE.2018.2798605}}.

\bibitem{schneier1999attack}
B.~Schneier, Attack trees, Dr. Dobb’s journal 24~(12) (1999) 21--29.

\bibitem{orojloo_evaluating_2015}
H.~Orojloo, M.~A. Azgomi, Evaluating the complexity and impacts of attacks on
  cyber-physical systems, in: 2015 {CSI} {Symposium} on {Real}-{Time} and
  {Embedded} {Systems} and {Technologies} ({RTEST}), IEEE, 2015, pp. 1--8.
\newblock \href {https://doi.org/10.1109/RTEST.2015.7369840}
  {\path{doi:10.1109/RTEST.2015.7369840}}.

\bibitem{ingoldsby2010attack}
T.~R. Ingoldsby, Attack tree-based threat risk analysis, Amenaza Technologies
  Limited (2010) 3--9.

\bibitem{meland_seamonster_nodate}
P.~H. Meland, D.~G. Spampinato, E.~Hagen, E.~T. Baadshaug, K.-m. Krister, K.~S.
  Velle, {SeaMonster}: {Providing} tool support for security modeling.

\bibitem{edge_using_2006}
K.~S. Edge, G.~C. Dalton, R.~A. Raines, R.~F. Mills, Using {Attack} and
  {Protection} {Trees} to {Analyze} {Threats} and {Defenses} to {Homeland}
  {Security}, in: {MILCOM} 2006 - 2006 {IEEE} {Military} {Communications}
  conference, 2006, pp. 1--7, iSSN: 2155-7586.
\newblock \href {https://doi.org/10.1109/MILCOM.2006.302512}
  {\path{doi:10.1109/MILCOM.2006.302512}}.

\bibitem{cornford_towards_2003}
S.~Cornford, T.~Paulos, L.~Meshkat, M.~Feather,
  \href{http://ieeexplore.ieee.org/document/1235492/}{Towards more accurate
  life cycle risk management through integration of ddp and pra}, in: 2003
  {IEEE} {Aerospace} {Conference} {Proceedings} ({Cat}. {No}.{03TH8652}),
  Vol.~2, IEEE, Big Sky, Montana, USA, 2003, pp. 2\_807--2\_814.
\newblock \href {https://doi.org/10.1109/AERO.2003.1235492}
  {\path{doi:10.1109/AERO.2003.1235492}}.
\newline\urlprefix\url{http://ieeexplore.ieee.org/document/1235492/}

\bibitem{feather_quantitative_2003}
M.~S. Feather, S.~L. Cornford,
  \href{http://link.springer.com/10.1007/s00766-002-0160-y}{Quantitative
  risk-based requirements reasoning}, Requirements Engineering 8~(4) (2003)
  248--265.
\newblock \href {https://doi.org/10.1007/s00766-002-0160-y}
  {\path{doi:10.1007/s00766-002-0160-y}}.
\newline\urlprefix\url{http://link.springer.com/10.1007/s00766-002-0160-y}

\bibitem{NIST_2020}
{Joint Task Force Interagency Working Group},
  \href{https://nvlpubs.nist.gov/nistpubs/SpecialPublications/NIST.SP.800-53r5.pdf}{Security
  and {Privacy} {Controls} for {Information} {Systems} and {Organizations}},
  Tech. rep., National Institute of Standards and Technology, edition: Revision
  5 (Sep. 2020).
\newblock \href {https://doi.org/10.6028/NIST.SP.800-53r5}
  {\path{doi:10.6028/NIST.SP.800-53r5}}.
\newline\urlprefix\url{https://nvlpubs.nist.gov/nistpubs/SpecialPublications/NIST.SP.800-53r5.pdf}

\bibitem{peddoju_file_2020}
S.~K. Peddoju, H.~Upadhyay, L.~Lagos,
  \href{https://onlinelibrary.wiley.com/doi/abs/10.1002/cpe.5825}{File
  integrity monitoring tools: {Issues}, challenges, and solutions}, Concurrency
  and Computation: Practice and Experience 32~(22) (2020) e5825, \_eprint:
  https://onlinelibrary.wiley.com/doi/pdf/10.1002/cpe.5825.
\newblock \href {https://doi.org/10.1002/cpe.5825}
  {\path{doi:10.1002/cpe.5825}}.
\newline\urlprefix\url{https://onlinelibrary.wiley.com/doi/abs/10.1002/cpe.5825}

\end{thebibliography}

\end{document}